\documentclass[aps,prb,preprint,superscriptaddress]{revtex4-1}

\usepackage{graphicx,amsmath,amsfonts}
\usepackage{epstopdf}

\begin{document}


\title{Double-exchange theory of ferroelectric polarization in orthorhombic manganites
with twofold periodic magnetic texture}


\author{I. V. Solovyev}
\email{SOLOVYEV.Igor@nims.go.jp}
\affiliation{Computational Materials Science Unit,
National Institute for Materials Science, 1-2-1 Sengen, Tsukuba,
Ibaraki 305-0047, Japan}
\affiliation{
Department of Theoretical Physics and Applied Mathematics, Ural Federal University,
Mira str. 19, 620002 Ekaterinburg, Russia}
\author{S. A. Nikolaev}
\affiliation{
Department of Theoretical Physics and Applied Mathematics, Ural Federal University,
Mira str. 19, 620002 Ekaterinburg, Russia}


\date{\today}

\begin{abstract}
We argue that many aspects of
improper ferroelectric activity in manganites with the $Pbnm$ and $P2_1nm$
orthorhombic structure
can be
rationalized by considering the limit of infinite intra-atomic splitting between the
majority- and minority-spin states (or the double exchange limit), which reduces the problem to
the analysis of a spinless double exchange (DE) Hamiltonian.
We apply this strategy to the low-energy model,
derived from the first-principles electronic structure calculations,
and combine it with the Berry-phase
theory of electric polarization.
We start with the simplest two-orbital model, describing the behavior of the $e_g$ bands,
and apply it to the
$E$-type antiferromagnetic (AFM) phase, which
in the DE limit
effectively breaks up into one-dimensional zigzag chains. We derive an analytical
expression for the electronic polarization (${\bf P}^{\rm el}$) and
explain how it depends on the orbital ordering and the energy
splitting $\Delta$ between $e_g$ states. Then, we evaluate parameters of this model for the
series of manganites. For these purposes we start from a more general five-orbital model
for all Mn $3d$ bands
and construct a new downfolded model for the $e_g$ bands.
From the analysis of these parameters, we conclude that the behavior of
${\bf P}^{\rm el}$ in
realistic manganites always corresponds to the limit of large $\Delta$. This property holds
for all considered compounds
even in the local-density approximation, which typically underestimates $\Delta$.
We further utilize this property in order to derive an analytical expression for ${\bf P}^{\rm el}$
in a general two-fold periodic magnetic texture, based on
the five-orbital model and
the perturbation-theory expansion for the Wannier functions in the first order of $1/\Delta$.
This expression explains the functional dependence of ${\bf P}^{\rm el}$ on the
relative directions of spins.
Furthermore, it suggests that ${\bf P}^{\rm el}$
is related to the asymmetry of
the transfer integrals, which should
simultaneously have symmetric and antisymmetric
components. Finally, we explain how
the polarization can be switched between
orthorhombic directions
$\boldsymbol{a}$ and $\boldsymbol{c}$
by inverting the zigzag AFM texture in every second
$\boldsymbol{ab}$ plane. We argue that this property is generic
and can be realized even in the twofold periodic texture.
\end{abstract}


\maketitle

\section{\label{sec:Intro} Introduction}

  The multiferroic materials (or multiferroics), where ferroelectricity coexists with some
long-range magnetic order, have attracted a great deal of attention.\cite{MF_review}
A very special class of multiferroics is improper ferroelectrics. In the latter case, the
ferroelectric (FE) polarization not only coexists, but can be induced by the magnetic order.
The improper ferroelectrics are expected to display a strong magneto-electric
coupling, which is extremely important for practical applications. For instance, because of
such coupling, the FE polarization can be efficiently controlled by the magnetic field, while
the magnetization can be controlled by the electric field.
From a technological point of view, the ultimate goal is to
find materials with the large FE polarization, which would be coupled to the magnetic texture
at maximally possible temperature (meaning that the magnetic transition temperature
should be also high).

  Manganites, crystalizing in the orthorhombic $Pbnm$ and $P2_1nm$ structure,
are regarded as one of the key multiferroic materials.
Despite low magnetic transition temperature (typically, less than 40 K) and modest values of the
FE polarization (less than $1$ $\mu$C/cm$^2$), which have been achieved so far,\cite{Ishiwata}
they have all essential ingredients to be called improper ferroelectrics. Namely, the appearance of ferroelectricity
coincides with some complex magnetic ordering. Moreover, the possibility of
switching the electric polarization by the magnetic fields has been directly demonstrated
experimentally.\cite{Kimura}
Therefore, these materials are fundamentally important and are typically used as a playground for
testing various theories and models of multiferroicity.

  Nevertheless, the theoretical understanding of improper ferroelectricity in
these compounds is still rather controversial and there is no unique view on the origin of this effect.
First, all multiferroic manganites are rather artificially divided in two groups:
\begin{itemize}
\item[(i)]
the systems
with the twofold periodic $E$-type antiferromagnetic (AFM) texture (such as HoMnO$_3$ and YMnO$_3$),
where the FE activity is attributed
to the nonrelativistic exchange striction,\cite{SergienkoPRL,Picozzi} and
\item[(ii)]
the rest of the systems, with more general magnetic periodicity, where the FE activity is believed to be
due to the relativistic spin-orbit (SO) interaction and the magnetic texture itself is ascribed
to the spin spiral.\cite{spiral_theories} The typical example of such systems is TbMnO$_3$, which has nearly
fourfold periodic magnetic texture.
\end{itemize}
This point was rationalized in the previous publications of one of the authors (Ref.~\onlinecite{PRB11,PRB12}),
where it was argued that there is no conceptual difference between twofold periodic and other
multiferroic manganiets. The relativistic SO interaction plays an equally important role in both cases:
as it deforms the $E$-type AFM state in the direction of the spin spiral,
it will also deform the spin spiral and form a more general spatially inhomogeneous magnetic state.
Thus, the ground state of multiferroic manganites will be neither the collinear $E$-state nor
the homogeneous spin spiral. The relativistic SO interaction is essential for producing this inhomogeneity.
However, the FE polarization itself is a nonrelativistic quantity in the sense that, for a given
inhomogeneous distribution of spins, the appearance of the FE polarization can be described
by nonrelativistic theories.

  Another group of controversies is related to the question: How to calculate the polarization and what
is the main contribution to it? Most of model calculations rely on the purely ionic picture, where the
noncentrosymmetric distribution of spins gives rise to noncentrosymmetric atomic displacements.
Then, the polarization is evaluated
in the framework of the point charge model.\cite{SergienkoPRL,Mochizuki} On the other hand, all modern first-principles
calculations of the FE polarization are based on the Berry-phase theory.\cite{KSV,Resta}
Besides ionic polarization, the Berry-phase theory prescribes the existence of an electronic term.
The latter can be expressed through the Wannier functions and is reduced to the ionic polarization only if
the Wannier functions are fully localized at the atomic sites. In this sense, the deviation from the
ionic picture is a measure of itineracy of the system. Moreover, unlike the ionic contribution,
the electronic polarization can be finite even in the centrosymmetric crystal structure, provided that
the inversion symmetry is broken by a magnetic order. Thus, the Berry-phase theory
excellently suits for improper ferroelectrics. The first-principles calculations
show that the electronic polarization can be as large as or even exceed the ionic contribution.\cite{Picozzi}
Nevertheless, the physical meaning of this effect is still rather obscure and
the electronic polarization is largely ignored in
model calculations
of multiferroic manganites.

  The purpose of this work is to make a bridge between first-principle electronic structure
calculations and models of the FE polarization.
Our main message is that the electronic polarization is important and
cannot be ignored. In the model calculations, it can be described by some ``superexchange type''
theories, similar to interatomic magnetic interactions.\cite{PWA,KugelKhomskii}
On the other hand, in the first-principles calculations, one should pay a special attention to the
relative direction of the electronic and ionic polarization: because of additional approximations,
results of theoretical structural optimization do not necessarily guarantee the correct answer to this question.

  Our analysis will be based on results of two previous works (Refs.~\onlinecite{PRB11,PRB12}), where
\begin{itemize}
\item[(i)]
A realistic low-energy model for the Mn $3d$ bands of manganites was constructed on the basis of first-principles
electronic structure calculations in the local-density approximation (LDA);
\item[(ii)]
This model was applied for the search of the magnetic ground state of orthorhombic manganites;
\item[(iii)]
The model calculations were supplemented with the
Berry-phase theory for the analysis of the FE polarization and its dependence on
the form of the magnetic ground state.
\end{itemize}
In this work we will further rationalize the story. First, we will show that the behavior of the
FE polarization can be well described in the framework of the
double exchange (DE) theory.\cite{DE_oldies}
The definition of the DE Hamiltonian will be given in Sec.~\ref{sec:Idea}. Particularly,
we will show that with the proper definition of the DE model, which should include effects of
orbital polarization of Coulombic origin, one can reproduce, even quantitatively, the values
of FE polarization obtained in a more general mean-field Hartree-Fock (HF) calculations for the low-energy model.
Then, we will introduce an analytically solvable model for the
$e_g$ electrons in the single zigzag chain (Sec.~\ref{subsec:analytics}) and argue that,
besides double exchange, the behavior of electronic polarization in realistic manganites
always corresponds to the limit of large intra-atomic energy splitting $\Delta$ between $e_g$ states
(Sec.~\ref{subsec:parameters}). It will allow us to further generalize our story and
derive an analytical expression for the
electronic polarization in an arbitrary twofold periodic magnetic texture,
based on the perturbation theory expansion for the Wannier functions
in the first order of $1/\Delta$ (Sec.~\ref{subsec:5orbitals}). The idea itself has some similarities with the superexchange
theory of interatomic magnetic interactions.\cite{PWA,KugelKhomskii} This analytical expression
nicely explains the behavior of electronic polarization in the low-energy model as well as in the
more general first-principles calculations. It also provides a good quantitative estimate for the polarization.
In Sec.~\ref{subsec:problems}, we will present a critical analysis of relative directions of electronic and ionic polarizations
in the experimental and theoretically optimized $P2_1nm$ structures of YMnO$_3$.
Then, in Sec.~\ref{subsec:switching},
we will explain how the electronic polarization can be manipulated by changing the magnetic texture.
Finally, in Sec.~\ref{sec:summary}, we draw our conclusions.

\section{\label{sec:Idea} Basic Idea and Approximations}

  The starting point of our work is that the main electronic and magnetic
properties of multiferroic manganites can be described reasonably well by the
one-electron
Hamiltonian:
\begin{equation}
\hat{H}^{\rm MF}_{ij} = \hat{t}_{ij} + \hat{\cal V}_i \delta_{ij},
\label{eqn:MF}
\end{equation}
which is constructed in the basis of Wannier orbitals for the Mn $3d$ bands.
In this notations, the matrix $\hat{t}_{ij}$ has site-diagonal ($i=j$) and off-diagonal ($i \ne j$) elements:
the former describes the crystal-field effects, while the latter stands for transfer integrals.
We do not consider explicitly the relativistic spin-orbit (SO) interaction. More specifically, it is assumed that the
SO interaction is important for specifying the directions of spins in some noncollinear
magnetic texture. However, it is unimportant for calculations of the FE polarization itself,
provided that the directions of spins are known and
the corresponding magnetic texture can be described by
appropriate rotations of the mean-field potentials $\hat{\cal V}_i$,
which will be specified below. Therefore, the matrix
$\hat{t}_{ij}$ does not depend on the spin-indices,
$s(s')$$=$ $\uparrow$ or $\downarrow$,
and can be presented in the form $\hat{t}_{ij} = \| t_{ij}^{mm'} \delta_{ss'}  \|$. In the more general
five-orbital model,
that we consider,
the indices $m$ and $m'$ have the following order:
$m(m')$$=$ $xy$, $yz$, $3z^2$$-$$r^2$, $zx$, or $x^2$$-$$y^2$. In the
two-orbital model, constructed only for the $e_g$ bands, the indices $m$ and $m'$
run over $3z^2$$-$$r^2$ and $x^2$$-$$y^2$.
$\hat{\cal V}_i$ in Eq.~(\ref{eqn:MF}) is the self-consistent one-electron potential, which is
constructed using parameters of effective Coulomb interactions and the density matrix for the Mn $3d$ states.
Generally, $\hat{\cal V}_i$ depends on both spin and orbital indices.

  In practice, the electronic low-energy model can be derived from the first-principles electronic structure
calculations, starting from the local-density approximation (LDA).\cite{review2008} The construction of the model can be
formulated rather rigorously in the basis of Wannier orbitals
for the Mn $3d$ bands. Then, $\hat{t}_{ij}$ is identified with the matrix elements of the
LDA Hamiltonian in the Wannier basis. Thus, without $\hat{\cal V}_i$, the parameters $\hat{t}_{ij}$ describe
the LDA electronic structure for the Mn $3d$ bands. The parameters of effective Coulomb
interactions for the Mn $3d$ bands can be derived, also in the Wannier basis, using
constrained random-phase approximation and/or the constrained LDA approach. For details, the reader is referred to the
review article (Ref.~\onlinecite{review2008}). Then, the model can be solved in the mean-field HF
approximation, which gives us the potentials $\hat{\cal V}_i$.\cite{review2008}

  After the solution, the FE polarization
can be obtained by applying the Berry-phase theory.\cite{KSV,Resta} Namely, the FE polarization
is divided into the ionic (${\rm ion}$) and electronic (${\rm el}$) parts:
$$
{\bf P} = {\bf P}^{\rm ion} + {\bf P}^{\rm el}.
$$
The ionic term reflects the non-cenrosymmetricity of the crystal structure itself
and is associated with the
displacements ($\Delta \boldsymbol{\tau}_i$) of ionic charges ($Z_i$) away from the centrosymmetric positions:
\begin{equation}
{\bf P}^{\rm ion} = \frac{1}{V}\sum_i Z_i \Delta \boldsymbol{\tau}_i,
\label{eqn:Pion}
\end{equation}
where $V$ is the primitive cell volume.
The electronic term reflects the fact of the inversion symmetry breaking in the
form of the wavefunctions, obtained
from the solution of quantum-mechanical Schr\"odinger equations. It incorporates the
effects of the magnetic inversion symmetry breaking and can take place even for
centrosymmetric crystalline systems, provided that the inversion symmetry is broken
by magnetic or some other electronic degrees of freedom.
The electronic term can be
computed in the reciprocal space,
by using the formula of King-Smith and Vanderbilt:\cite{KSV}
\begin{equation}
{\bf P}^{\rm el} = - \frac{ie}{(2 \pi)^3} \sum_{n = 1}^M
\int_{\rm BZ} \langle n {\bf k} | \nabla_{\bf k} | n {\bf k} \rangle d {\bf k} ,
\label{eqn:PKSV}
\end{equation}
where $| n {\bf k} \rangle$ is the cell periodic wavefunction, the summation runs over the
occupied bands ($n$), the ${\bf k}$-space integration goes over the first
Brillouin zone, and $-$$e$ ($e > 0$) is the electron charge.
In practical calculations, Eq.~(\ref{eqn:PKSV}) is replaced by a discrete grid formula.\cite{Resta}
Eq.~(\ref{eqn:PKSV}) can be also rewritten in terms of the Wannier function ($w_n$), constructed from
$| n {\bf k} \rangle$ in the real space:\cite{KSV}
\begin{equation}
{\bf P}^{\rm el} = - \frac{e}{V} \sum_{n = 1}^M
\int {\bf r} | w_n({\bf r}) | d {\bf r}.
\label{eqn:PW}
\end{equation}
In all these equations, it is understood that ${\bf P}$ is the \textit{change} of the polarization,
obtained in the process of adiabatic lowering of the inversion symmetry.\cite{Resta}
Moreover, the contribution of the low-energy bands (in our case, the Mn $3d$ bands) is
accounted by ${\bf P}^{\rm el}$. Therefore, the contribution of all other occupied states,
which are not included to the low-energy model, should be described (at least, approximately)
by ${\bf P}^{\rm ion}$. Then, since the oxygen $2p$ band is fully occupied,
it is reasonable to take $Z_{\rm O} = -$$2e$, which corresponds to the formal valence state of O$^{2-}$.
On the other hand, all valence states of the rare-earth (RE) ions are empty.
This should correspond to $Z_{\rm RE} = 3e$. In the noncentrosymmetric $P2_1nm$ structure,
the Mn sites
do not contribute to ${\bf P}^{\rm ion}$.\cite{PRB12} Therefore, the parameter
$Z_{\rm Mn}$ is not important for our purposes.

  In the previous publications, this procedure was applied to the series of orthorhombic manganites.
Particularly, the behavior of parameters of the low-energy model, derived from the
first-principles electronic structure calculations, was discussed in Ref.~\onlinecite{JPSJ}.
An example of such parameters for YMnO$_3$ can be found in Supplemental Material of Ref.~\onlinecite{PRB12}.
The properties of the magnetic ground state, obtained from the
solution of the low-energy model in the HF approximation, and corresponding behavior of the FE polarization
were considered in Refs.~\onlinecite{PRB11,PRB12}. Note that a scaling factor was missing in the
calculations of the FE polarization reported in Ref.~\onlinecite{PRB11}. This error was corrected in Ref.~\onlinecite{PRB12}.

  As far as the FE polarization is concerned,
the low-energy model reproduces results of the first-principles electronic structure
calculations
(Refs.~\onlinecite{Picozzi,Okuyama,Yamauchi})
on a good semi-quantitative level. Moreover, the low-energy model was very helpful in clarifying
details of the noncollinear magnetic ground state, which can be realized in orthorhombic manganites,
namely: (i) the canting of spins and magnetic origin of the
twofold periodic phase;\cite{PRB11,PRB12}
(ii) deformation of the spin-spiral texture, yielding FE activity in both two- and fourfold periodic systems;\cite{PRB11}
(iii) the absence of the magnetic inversion symmetry breaking in systems with odd magnetic periodicity.\cite{PRB11}

  In this work, we will further rationalize the story by considering the DE limit for the
FE polarization.

  Let us start with the ferromagnetic (FM) state, where each $\hat{\cal V}_i$ is diagonal
with respect to the spin indices,
$$
\hat{\cal V}_i =
\left(
\begin{array}{cc}
\hat{\cal V}_i^\uparrow & 0 \\
0 & \hat{\cal V}_i^\downarrow \\
\end{array}
\right),
$$
and $\hat{\cal V}_i^{\uparrow,\downarrow}$ are the $5$$\times$$5$ matrices in the orbital subspace.
The states with $s = \uparrow$ are occupied by four electrons and the ones with $s = \downarrow$
are empty. Then, $\hat{\cal V}_i^{\downarrow}$ can be identically presented in the form:
$\hat{\cal V}_i^{\downarrow} = \Delta_{\rm ex} + \Delta \hat{\cal V}_i^{\downarrow}$, where
$\Delta_{\rm ex}$ is the intra-atomic exchange splitting between centers of gravity of the
majority ($\uparrow$) and minority ($\downarrow$) spin states, and $\Delta \hat{\cal V}_i^{\downarrow}$
describes the orbital splitting of unoccupied $\downarrow$-spin states. Moreover,
four $3d$ electrons obey Hund's first rule, which tend to
form the state with the maximal spin $S=2$.
Therefore, besides on-site Coulomb repulsion ($U$), $\Delta_{\rm ex}$ will contain a large
contribution, being proportional to the local magnetic moment
($2S$) and the intra-atomic exchange coupling ($J_{\rm H}$).
This is the main reason why
for many applications $\Delta_{\rm ex}$ can be treated as the largest physical parameter,
and the DE limit corresponds to the extreme situation where $\Delta_{\rm ex} \rightarrow \infty$.\cite{DE_oldies}
On the other hand, the splitting of unoccupied $\downarrow$-spin states
is considerably weaker. For example, in the HF approximation, it is caused by
relatively small nonsphericity of the
Coulomb potential.

  Therefore, when $\Delta_{\rm ex} \rightarrow \infty$, the details
of (finite) splitting of the $\downarrow$-spin states become unimportant and
our first approximation is to replace
$\Delta \hat{\cal V}_i^{\downarrow}$
by $\hat{\cal V}_i^{\uparrow}$. It allows us to present $\hat{\cal V}_i$
in the following form:
\begin{equation}
\hat{\cal V}_i \approx \hat{\cal V}_i^\uparrow +
\left(
\begin{array}{cc}
0 & 0 \\
0 & \Delta_{\rm ex} \\
\end{array}
\right),
\label{eqn:Vapprox}
\end{equation}
where the orbital-dependent part ($\hat{\cal V}_i^\uparrow$) does not depend on the spin indices and
the spin-dependent part does not depend on the orbital ones. Therefore,
spin and orbital transformations of Eq.~(\ref{eqn:Vapprox}) can be
treated separately.

\begin{figure}
\begin{center}
\includegraphics[height=8cm]{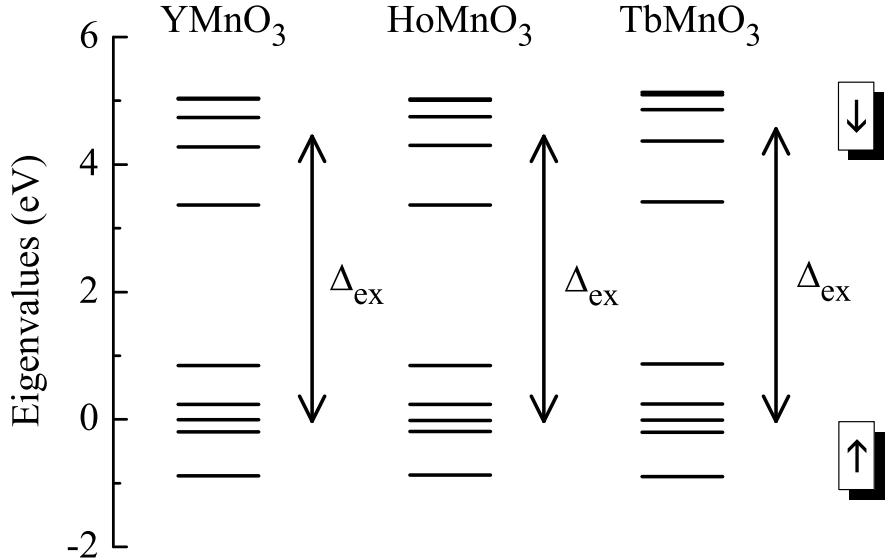}
\end{center}
\caption{\label{fig.potEV}
Eigenvalues of the Hartree-Fock potential, as obtained in the low-energy model
for the ferromagnetic phase
of YMnO$_3$, HoMnO$_3$, and TbMnO$_3$ (results of Refs.~\onlinecite{PRB11,PRB12} for the
experimental $Pbnm$ structure). $\Delta_{\rm ex}$ is the intra-atomic splitting
between centers of gravity of the majority ($\uparrow$) and minority ($\downarrow$)
spin states.
}
\end{figure}

  A typical example, illustrating the structure of the atomic $3d$ level splitting
by the Coulomb and exchange potentials
in the low-energy model, is shown in Fig.~\ref{fig.potEV}.
Typical values of $\Delta_{\rm ex}$ in manganites are about $4.5$ eV,
while the splitting of the $\downarrow$-spin states is about $1.7$ eV.
The difference is not extremely large.
However, as we will see in a moment, it is sufficient to justify the use of the DE limit
for the FE polarization.

  As the next step,
let us consider an arbitrary magnetic texture, where the directions of spin (${\bf e}_i$)
at each site of the lattice
are specified by the combinations of polar ($\theta_i$) and azimuthal ($\phi_i$) angles:
${\bf e}_i = (\cos \phi_i \sin \theta_i, \sin \phi_i \sin \theta_i, \cos \theta_i)$.
Corresponding electronic structure can be generated by the unitary transformation of Eq.~(\ref{eqn:Vapprox}), using
spin-rotation matrices:
\begin{equation}
\hat{\cal V}_i \rightarrow \hat{U}(\theta_i, \phi_i) \hat{\cal V}_i \hat{U}^\dagger(\theta_i, \phi_i),
\label{eqn:Rpot}
\end{equation}
where
$$
\hat{U}(\theta_i, \phi_i) =
\left(
\begin{array}{cc}
\cos \frac{\theta_i}{2} & \sin \frac{\theta_i}{2} e^{-i \phi_i} \\
-\sin \frac{\theta_i}{2} e^{i \phi_i} & \cos \frac{\theta_i}{2} \\
\end{array}
\right).
$$
Here, it is assumed that the angles $(\theta_i, \phi_i)$ are specified by magnetic interactions in the system
(for the form of the optimized magnetic textures, the reader is referred to Refs.~\onlinecite{PRB11,PRB12})
and the one-electron potential for an arbitrary direction of spin can be obtained by using
rigid spin rotations [Eq.~(\ref{eqn:Rpot})]
without additional self-consistency. This is a very good approximation in the case of manganites, because:
\begin{itemize}
\item[(i)]
Due to the strong Hund's coupling, the local spin magnetization will always tend to stay in the
saturated state. Therefore, the absolute value of this magnetization will only weakly depend on the direction
of spins at other magnetic sites.
\item[(ii)]
The orbital configuration is rigidly fixed by the Jahn-Teller (JT) distortion and practically
does not depend on the type of the spin texture. For example, the energy splitting of the
$e_g$ states, caused by the JT distortion, is about $1.5$ eV, while typical strength of
interatomic exchange interactions is of the order of several meV.\cite{JPSJ}
The exchange interactions can be additionally optimized by means of the orbital reconstruction,
which works against the JT splitting.\cite{KugelKhomskii}
However, the possible energy gain, caused by this reconstruction (typically, of the order of the exchange interactions
themselves) is much smaller than the energy of the JT distortion. Thus, the orbital
reconstruction does not occur.
\end{itemize}

  The next step is to transform Eq.~(\ref{eqn:Rpot}) to the local coordinate frame,
corresponding to the $z$ direction of magnetization at each site of the lattice. It leads to the
following transformation of the transfer integrals:
$$
\hat{t}_{ij} \rightarrow \hat{U}^\dagger(\theta_i, \phi_i) \hat{t}_{ij} \hat{U}(\theta_j, \phi_j).
$$
Then, taking the limit $\Delta_{\rm ex} \rightarrow \infty$, we obtain the well known DE model:
\begin{equation}
\hat{H}^{\rm DE}_{ij} = \xi_{ij} \hat{t}_{ij} + \hat{\cal V}_i^\uparrow \delta_{ij},
\label{eqn:DE}
\end{equation}
which formulated in the subspace of the $\uparrow$-spin states, in the local coordinate frame.\cite{DE_oldies}
The prefactor $\xi_{ij}$ is nothing but the $\uparrow \uparrow$-element of the product
$\hat{U}^\dagger(\theta_i, \phi_i) \hat{U}(\theta_j, \phi_j)$:
$$
\xi_{ij} = \cos \frac{\theta_i}{2} \cos \frac{\theta_j}{2} +
\sin \frac{\theta_i}{2} \sin \frac{\theta_j}{2} e^{-(\phi_i - \phi_j)},
$$
which satisfies the well known property: $\xi_{ij}$$=$ $1$ and $0$ for the ferromagnetically and antiferromagnetically
coupled spins, respectively. Therefore, in the DE limit, any antiferromagnetic (AFM) phase
effectively breaks up into FM segments. For example,
the description of the $E$-type AFM phase is reduced to the analysis of
one-dimensional FM zigzag chains.\cite{SergienkoPRL,Barone}

\begin{figure}
\begin{center}
\includegraphics[height=10cm]{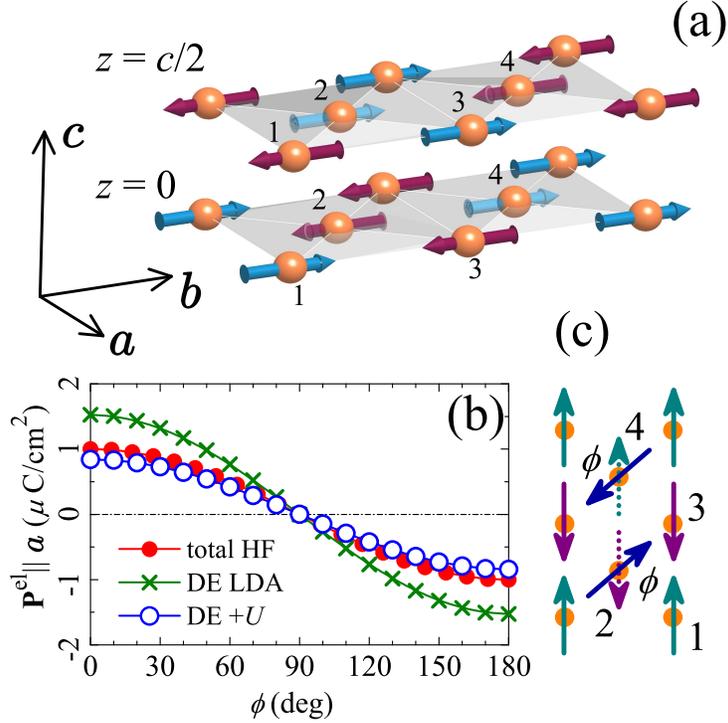}
\end{center}
\caption{\label{fig.Ecanting}(Color online)
(a) $E$-type antiferromagnetic texture. (b) Behavior of electronic polarization in YMnO$_3$ upon
rotation of magnetic moments as obtained in the self-consistent mean-field Hartree-Fock approximation
(total HF); in the double exchange model for the LDA band structure (DE LDA); and in the
double exchange model with the Hartree-Fock potential $\hat{\cal V}_i^\uparrow$
(DE $+$$U$). In the rotated texture, the directions of spins
at the sites $1$ and $3$ were fixed, while the spins at the sites $2$ and $4$ were
rotated by the angle $\phi$, as explained in panel (c). The planes
$z=0$ and $z=c/2$ were coupled antiferromagnetically.
}
\end{figure}

  Next, we investigate abilities of the DE model for the description of the FE polarization.
For these purposes, we calculate the electronic structure for the DE Hamiltonian [Eq.~(\ref{eqn:DE})],
and then evaluate the electronic polarization, using the Berry-phase formula [the discrete analog of Eq.~(\ref{eqn:PKSV})].\cite{KSV,Resta}
This procedure was applied to the series of orthorhombic manganites
TbMnO$_3$, HoMnO$_3$, and YMnO$_3$ (and using both experimental and theoretically
optimized crystal structure for the latter compound).\cite{PRB11,PRB12}
The obtained polarization was compared with results of self-consistent HF calculations for the
same low-energy model, but without additional approximations associated with the use of the DE limit.
Typical results of such calculations are illustrated in Fig.~\ref{fig.Ecanting} for the $Pbnm$ phase of YMnO$_3$
(other systems show very similar behavior).
More specifically, we consider a twofold periodic magnetic texture, which is
explained in Fig.~\ref{fig.Ecanting}(c), and
keep the AFM coupling between adjacent planes $z=0$ and $z=c/2$, as explained in Fig.~\ref{fig.Ecanting}(a).
Then, $\phi =$ $0$ and $180^\circ$ correspond to the
AFM alignment of the $E$-type, while $\phi =$ $90^\circ$ corresponds to the spin-spiral alignment.
For this geometry, the FE polarization should be parallel to the orthorhombic $\boldsymbol{a}$ axis.\cite{Picozzi}
In the DE model itself, we consider two levels of approximations. In the first case (denoted as `DE LDA'),
we neglect
$\hat{\cal V}_i^\uparrow$ and consider only the
crystal-field splitting and transfer integrals, derived from the LDA band structure.
Then, the transfer integrals are modulated by $\xi_{ij}$, as requested by the DE model.
In the second case, we consider the full DE Hamiltonian, Eq.~(\ref{eqn:DE}), including $\hat{\cal V}_i^\uparrow$
(denoted as `DE $+$$U$'). All magnetic solutions are insulating. Therefore, we can use the
Berry-phase formula for the analysis of ${\bf P}^{\rm el}$.
The DE LDA scheme overestimates
the electronic polarization by about 50 \%. Nevertheless, this is to be expected, because LDA underestimates the
band gap. Therefore, the FE polarization should be generally larger. Similar behavior was
found in the first-principles calculations.\cite{Picozzi,Okuyama} The analytical expression, explaining the
band-gap dependence of ${\bf P}^{\rm el}$, will be derived in the next section.
The band-gap problem is corrected by $\hat{\cal V}_i^\uparrow$. Therefore, the FE polarization, derived in the
DE $+$$U$ scheme, is smaller. Moreover, results of self-consistent
HF calculations for the electronic polarization are well reproduced by the DE $+$$U$ scheme:
although
${\bf P}^{\rm el}$ in the approximate
DE $+$$U$ scheme is systematically smaller,
the typical difference, which was obtained for all considered systems, is less than 15 \%.

  This is our main observation and also the main motivation of the rest of our work. By considering the DE limit,
we will slightly lose in the accuracy. But instead we will be able to rationalize the problem and derive
several analytical expressions for the FE polarization in orthorhombic manganites. Our analysis will also clarify
results of the low-energy model and
first-principles calculations.

\section{\label{sec:Results} Results}

  We start with the analysis of the $E$-type AFM phase. As was pointed out above, in
the DE limit, the FE AFM $E$-phase breaks up into one-dimensional FM zigzag chains.
Therefore, the key moment for understanding the origin of the FE activity in the $E$-phase is the
analysis of isolated zigzag chain.\cite{Barone} In Sec.~\ref{subsec:analytics}, we start such an analysis
with the simplest but analytically solvable model for the $e_g$ electrons.
In Sec.~\ref{subsec:parameters} we will derive parameters of such a model, starting from a more general five-orbital model,
which was obtained from the first-principles calculations.\cite{JPSJ,PRB11,PRB12}
From the analysis of this model we will conclude that
the situation, realized in most of the electronic structure calculations (even in ordinary LDA),
corresponds to the limit of
large energy splitting $\Delta$ between atomic $e_g$ states, which incorporates the
effects of the JT distortion and (optionally) the on-site Coulomb repulsion. Then, by considering the large-$\Delta$ limit,
in Sec.~\ref{subsec:5orbitals}
we will derive an analytical expression for the FE polarization, which is based on the five-orbital model.
This expression explains
the functional dependence of ${\bf P}^{\rm el}$ on the relative directions of spins and the form of
nearest-neighbor transfer integrals.
In Sec.~\ref{subsec:problems} we will analyze relative directions of
electronic and ionic polarizations
in the noncentrosymmetric $P2_1nm$ structure and point out on the problem of structural optimization,
which apparently exists in some of the first-principles calculations, where the directions of noncentrosymmetric
atomic displacements are inconsistent with the type of the orbital ordering, realized in the
FM zigzag chain. In Sec.~\ref{subsec:switching}, we discuss the possibility of switching the FE polarization
by changing the magnetic texture: we argue that,
even in the twofold periodic texture, there is another type of the AFM zigzag ordering, which leads to a finite
FE polarization along the orthorhombic $\boldsymbol{c}$ axis. However, the value of this polarization
is expected to be small.

\subsection{\label{subsec:analytics} Analytically solvable model for the
$e_g$ electrons in the zigzag chain}

\begin{figure}
\begin{center}
\includegraphics[height=5cm]{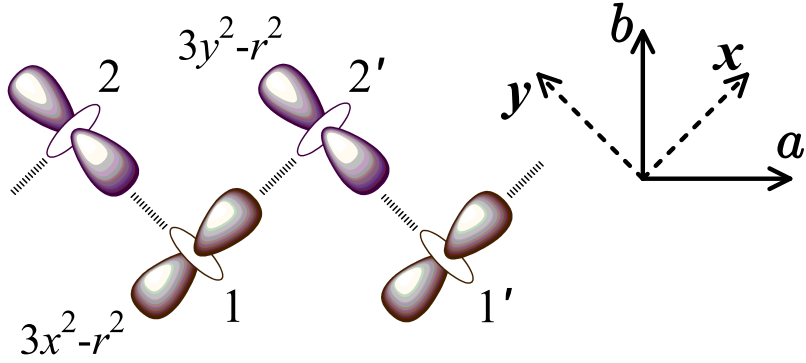}
\end{center}
\caption{\label{fig.ModelChain}(Color online)
Geometry of the zigzag chain for the square lattice and the occupied $e_g$ orbitals of the
$3x^2$$-$$r^2$ and $3y^2$$-$$r^2$ type.
Cubic and orthorhombic axes are denoted as $\boldsymbol{xy}$ and $\boldsymbol{ab}$, respectively.
}
\end{figure}

  The zigzag chain consists of the two groups of sites: the lower corner sites $1$ and the
upper corner sites $2$ (see Fig.~\ref{fig.ModelChain}).
The orthorhombic translation $\boldsymbol{a}$ transforms each group to itself (the translated
sites are denoted as $1'$ and $2'$, respectively).
It is assumed that the lattice distortion stabilizes some $e_g$ orbitals
at the sites $1$ and $2$, which will be
denoted as $|1 \rangle_1$ and $|1 \rangle_2$, respectively.
The orthogonal to them $e_g$ orbitals are denoted as $|2 \rangle_1$ and $|2 \rangle_2$, respectively.
Furthermore, it is assumed that there is a symmetry operation ($\hat{S}$), which transforms the zigzag chain to itself and
which consists of the $180^\circ$ rotation around the $\boldsymbol{a}$ axis ($\hat{C}^2_a$) with consequent translation.
$\hat{S}$ will transform site $1$ to site $2$, and vice versa.
For the $Pbnm$ structure (and with some appropriate choice of the origin),
such symmetry operation is $\{ \hat{C}^2_a| \boldsymbol{a}/2$$+$$\boldsymbol{b}/2 \}$
(where the first part stands for the rotation, and the second part specifies the translation)
while for the $P2_1nm$ structure, it is
$\{ \hat{C}^2_a| \boldsymbol{a}/2$$+$$\boldsymbol{c}/2 \}$. It is important that
both symmetry operations include the translation $\boldsymbol{a}/2$.
Then, it is convenient to work in the local basis, corresponding
to the diagonal presentation of the $e_g$ level splitting, such that
$\hat{S}$ would transform the basis functions of the site $1$ to the ones of the site $2$, and vice versa.
Our idea is that, although we have two different sites, with such choice
of the basis functions, the
Hamiltonian becomes periodic with the period $\boldsymbol{a}/2$ and the problem can be treated as
if it would have only one site in the primitive cell. Similar idea was used for the analysis of the
CE AFM state in the half-doped manganites.\cite{PRB01}
Rather generally, these basis functions can be chosen in the form:
\begin{eqnarray}
|1\rangle_1 & = & -\cos \beta |3z^2-r^2\rangle_1 - \sin \beta |x^2-y^2\rangle_1, \label{eqn:L1} \\
|2\rangle_1 & = & \phantom{-} \sin \beta |3z^2-r^2\rangle_1 - \cos \beta |x^2-y^2\rangle_1 \label{eqn:L2}
\end{eqnarray}
at the site $1$, and
\begin{eqnarray}
|1\rangle_2 & = & -\cos \beta |3z^2-r^2\rangle_2 + \sin \beta |x^2-y^2\rangle_2, \label{eqn:L3} \\
|2\rangle_2 & = & \phantom{-} \sin \beta |3z^2-r^2\rangle_2 + \cos \beta |x^2-y^2\rangle_2 \label{eqn:L4}
\end{eqnarray}
at the site $2$, where $-$$\pi/2 < \beta \leq \pi/2$.
$| \beta | = 60^\circ$
corresponds to the ideal square lattice, subjected to the JT distortion.
Here, it is assumed that the direction of this distortion is determined by
anharmonic electron-lattice interactions, which stabilize orbitals of the type $| 1 \rangle$
with $| \beta |$ close to $60^\circ$.\cite{KugelKhomskii,Kanamori}
Then, all deformations of the orbital ordering pattern are described by the single parameter $\beta$.
Here, we continue to use the notations $|3z^2$$-$$r^2\rangle$ and $|x^2$$-$$y^2\rangle$ for the
$e_g$ orbitals, although it should be understood that they are valid only for the
ideal square lattice, and more generally we have in mind some $|3z^2$$-$$r^2\rangle$-like orbitals, which transform to
each other as $\hat{S}|3z^2$$-$$r^2\rangle_1 = |3z^2$$-$$r^2\rangle_2$, and some $|x^2$$-$$y^2\rangle$-like orbitals,
which transform to each other as $\hat{S}|x^2$$-$$y^2\rangle_1 = -$$|x^2$$-$$y^2\rangle_2$. Such deformations
of the ideal $e_g$ orbitals
can be caused, for example, by buckling distortions.
It is easy to check that $\beta = -$$60^\circ$ yields
$|1\rangle_1 = |3x^2$$-$$r^2 \rangle_1$, $|2\rangle_1 = |y^2$$-$$z^2 \rangle_1$,
$|1\rangle_2 = |3y^2$$-$$r^2 \rangle_2$, and $|2\rangle_2 = |x^2$$-$$z^2 \rangle_1$; while $\beta = 60^\circ$
yields $|1\rangle_1 = |3y^2$$-$$r^2 \rangle_1$, $|2\rangle_1 = |z^2$$-$$x^2 \rangle_1$,
$|1\rangle_2 = |3x^2$$-$$r^2 \rangle_2$, and $|2\rangle_2 = |z^2$$-$$y^2 \rangle_2$. Then, although in realistic
situations, $| \beta |$ can deviate from $60^\circ$, we will say that $\beta$$<$$0$ corresponds to the
$3x^2$$-$$r^2$/$3y^2$$-$$r^2$ type of the orbital ordering (referring to the
type of the occupied orbitals at the sites $1$/$2$),
while $\beta$$>$$0$ corresponds to the $3y^2$$-$$r^2$/$3x^2$$-$$r^2$ type of the orbital ordering.

  As for the transfer integrals between $e_g$ orbitals, we again consider a more general case and write them
in the following form:
\begin{equation}
\hat{t}_{12'} = - \frac{1}{2} \left(
\hat{\mathbb{I}} - | \sin \beta | \hat{\sigma}_x - \cos \beta \hat{\sigma}_z
\right),
\label{eqn:tright}
\end{equation}
for the bond $1$-$2'$, and
\begin{equation}
\hat{t}_{12} = - \frac{1}{2} \left(
\hat{\mathbb{I}} + | \sin \beta | \hat{\sigma}_x - \cos \beta \hat{\sigma}_z
\right),
\label{eqn:tleft}
\end{equation}
for the bond $1$-$2$,
in terms of the pseudospin Pauli matrices $\hat{\sigma}_x$, $\hat{\sigma}_y$, and $\hat{\sigma}_z$, and the
$2$$\times$$2$ identity matrix $\hat{\mathbb{I}}$. Throughout this section, all energies are in the units
of two-center integral $t_0$ of the $dd\sigma$ type.\cite{SK} The form of $\hat{t}_{ij}$
is suggested by
the $dd\sigma$ transfer integrals in the ideal square lattice, which again corresponds to
$ \beta  = 60^\circ$. Therefore, it is assumed that all deviations from the ideal
square lattice are described by the single parameter $\beta$, similar to the orbital ordering.
Note also that Eqs.~(\ref{eqn:tright}) and (\ref{eqn:tleft}) satisfy the idempotency condition
$(\hat{t}_{ij})^2 = \hat{t}_{ij}$, which holds for the $dd\sigma$ transfer integrals
in the square lattice.

  Thus, in our model, the orbital ordering and the transfer integrals are described by the same parameter $\beta$.
Generally speaking, these are different quantities, which should be specified by two different sets of parameters.
Nevertheless, in the analytical model, one would always like to reduce the number of independent parameters
to the minimum. Moreover, the use of the single parameter $\beta$ is indeed a very
reasonable approximation for our purposes:
\begin{itemize}
\item[(i)]
At least for the ideal square lattice, the orbital ordering and the transfer integrals can be
described by the same $| \beta | = 60^\circ$. Thus, there is the reference point where our
construction is exact;
\item[(ii)] Small deviations from the ideal case are treated as an approximation and we have some
freedom to decide the form of this approximation. In Sec.~\ref{subsec:parameters} we will show
that typical deviations of $| \beta |$ from $60^\circ$ are not large and, therefore, our
approximation is robust;
\item[(iii)] According to Eqs.~(\ref{eqn:tright}) and (\ref{eqn:tleft}), the transfer integrals
do not depend on the sign of $\beta$ (although the orbital ordering does). This is the very important requirement,
because the phases of transfer integrals is determined solely by the geometry of the zigzag chain and
should not depend on the type of the orbital ordering.
\end{itemize}

  After the transformation to the local basis, given by Eqs. (\ref{eqn:L1})-(\ref{eqn:L4}), the
transfer integrals become:
\begin{equation}
\hat{\mathsf{t}}_{12'} = \hat{\mathsf{t}}_{21} =
 \frac{1}{2} \left(
\cos \beta \hat{\mathbb{I}} + \sin 2\beta \hat{\sigma}_x - i| \sin \beta |  \hat{\sigma}_y - \cos 2 \beta \hat{\sigma}_z
\right),
\label{eqn:tlocal}
\end{equation}
where
$\hat{\mathsf{t}}_{2'1} = \hat{\mathsf{t}}_{12} = \hat{\mathsf{t}}_{21}^T$. Thus, the transfer integrals are indeed periodic with the
period $\boldsymbol{a}/2$ and, in the reciprocal space, the problem is reduced to the
analysis of the $2 \times 2$ Hamiltonian of the form:
$$
\hat{\cal H}(k) = \varepsilon (k) + {\bf d}(k) \cdot \hat{\boldsymbol{\sigma}},
$$
where
$\varepsilon (k) =  \cos \beta \cos (ka/2)$,
and components of the vector
${\bf d} \equiv (d_x,d_y,d_z)$ are given by
$d_x =  \sin 2 \beta \cos (ka/2)$, $d_y =  | \sin \beta | \sin (ka/2)$, and
$d_z = -$$\cos 2 \beta \cos (ka/2) - \Delta/2$. The parameter
$\Delta$ in $d_z$
is the intra-atomic energy splitting between $e_g$ states, caused by lattice distortions and
Coulomb interactions.
This result can be also viewed as if the transformation (\ref{eqn:L1})-(\ref{eqn:L4}) would ``straightened'' the
zigzag chain and made it equivalent to a linear chain, but with different transfer integrals operating in the
positive and negative directions of $\boldsymbol{a}$. Because of the condition $\hat{\mathsf{t}}_{12} = \hat{\mathsf{t}}_{21}^T$,
the transfer integrals are generally not centrosymmetric with respect to the atomic sites and the system
will develop a finite electronic polarization. Nevertheless, in the limit $\Delta$$\rightarrow$$\infty$, the
basis orbitals of the type `$2$' are projected out. Then, the transfer integrals
between orbitals of the same type `1' are just scalars, and the condition $\hat{\mathsf{t}}_{12} = \hat{\mathsf{t}}_{21}^T$
becomes equivalent to $\mathsf{t}_{12} = \mathsf{t}_{21}$. Thus, in the limit $\Delta$$\rightarrow$$\infty$, the
problem should become centrosymmetric. From this point of view, it is logical to consider
the limit $\Delta$$\rightarrow$$\infty$ as the reference point for the electronic polarization.

  The eigenvalues of $\hat{\cal H}(k)$ are given by
$E_\pm (k) = \varepsilon (k)$$\pm$$|{\bf d}(k)|$,
and the eigenvector, corresponding to the lowest occupied band, satisfies the
condition:
$[{\bf d}(k) \cdot \hat{\boldsymbol{\sigma}}$$+$$|{\bf d}(k)|] | -,k \rangle = 0$.\cite{Zhang}
Then, $| -,k \rangle$ can be taken in the form:
$$
| -, k \rangle =
\left(
\begin{array}{c}
C_1 (k) \\
e^{i \gamma(k)} C_2 (k)
\end{array}
\right),
$$
where
$$
C_1 (k) = \frac{1}{\sqrt{2}} \left(1 - \frac{d_z (k)}{|{\bf d} (k)|} \right)^{1/2},
$$
$$
C_2 (k) = -\frac{1}{\sqrt{2}} \left(1 + \frac{d_z (k)}{|{\bf d} (k)|} \right)^{1/2},
$$
and $\gamma(k) = \arctan (d_y/d_x)$.

  At the half-filling (one $e_g$ electron per each Mn site), the zigzag chain is a
band insulator. This property holds even for $\Delta = 0$ due to specific form
of the $dd \sigma$ transfer integrals.\cite{Hotta}
Moreover, the reciprocal lattice vector of the ``straightened'' chain is $G= 4\pi/a$, and
$| -,k \rangle$ is a periodic function of $G$. Therefore, the electronic polarization can be
computed directly, using the formula of King-Smith and Vanderbilt.\cite{KSV}
Note that in this section, it is more convenient to work with the electric dipole moment,
rather than with the polarization density. Therefore, Eq.~(\ref{eqn:PKSV}) was
additionally multiplied by the primitive cell volume $V$.
Nevertheless, unless it is specified otherwise, we will use the same notations for this
quantity and continue to call it ``the polarization''.
Then, we obtain the following expression for the FE polarization
parallel to the
orthorhombic $\boldsymbol{a}$ axis (per two Mn sites in the zigzag chain):
$$
P_E^{\rm el} = \frac{ea}{2 \pi} \int_{-2\pi/a}^{2\pi/a} C_2^2(k) \frac{d \gamma(k)}{d k} dk,
$$
which can be further transformed to
\begin{equation}
P_E^{\rm el} = \frac{ea^2}{4 \pi} \int_0^{2\pi/a}
\frac{| \sin \beta | \sin 2\beta}{|{\bf d}(k)|\left[ |{\bf d}(k)| - d_z(k) \right]} dk,
\label{eqn:Pfinal}
\end{equation}
where the subscript $E$ means that this polarization corresponds to the $E$-type AFM phase
in the DE limit.

  Thus, we immediately recognize that
when the orbital ordering changes from $3x^2$$-$$r^2$/$3y^2$$-$$r^2$ ($\beta$$<$$0$) to
$3y^2$$-$$r^2$/$3x^2$$-$$r^2$ ($\beta$$>$$0$), the polarization changes its sign.

  Then, it is straightforward to find that
$$
\lim_{\Delta \to 0^+} P_E^{\rm el} = \frac{| \sin \beta |}{\sin \beta} \frac{ea}{2}
$$
and, therefore, $|P_E^{\rm el}| = ea/2$ (see Ref.~\onlinecite{remark1}).
Then, since $P_E^{\rm el}$ is well defined modulo $ea$,\cite{KSV} the values of $P_E^{\rm el}$ and $-$$P_E^{\rm el}$ for $\Delta = 0$ are
equivalent. Such a situation means that
the system possesses the inversion symmetry, but the inversion centers are
located in the middles of the bonds.\cite{Zak}
Thus, by removing the JT distortion from our model, we effectively create a new
inversion center. This is indeed the case for the model considered above:
since $\hat{t}_{12} = \hat{t}_{21}$ and $\hat{t}_{12'} = \hat{t}_{2'1}$
[see Eqs.~(\ref{eqn:tright})-(\ref{eqn:tleft})], the transfer integrals are
centrosymmetric with respects to the middles of the bonds.

  In the limit $\Delta \rightarrow \infty$, we have
\begin{equation}
P_E^{\rm el}(\Delta \to \infty) \rightarrow \frac{ea  | \sin \beta | \sin 2 \beta}{\Delta^2}.
\label{eqn:Pinfinity}
\end{equation}
This result also has a transparent physical meaning and can be easily understood by starting
from the expression
\begin{equation}
P_E^{\rm el} = -2e \int x w^2(x) dx,
\label{eqn:PWannier}
\end{equation}
in terms of the Wannier functions,\cite{KSV} where the prefactor `2' stands for the number of Mn sites
in the primitive cell of the zigzag chain. Let us consider the limit $\Delta \rightarrow \infty$, where
$| w_{\infty}  \rangle = |1  \rangle_1$ and it is centered at the site $1$ (see Fig.~\ref{fig.ModelChain}).
Then, in the first order of $1/\Delta$, this Wannier function will have a finite tail, spreading to
the neighboring sites $2$ and $2'$, which are located at
$x = -$$a/2$ and $a/2$, respectively.
In the first order of perturbation theory,
this tail is proportional to the transfer integrals [Eq.~(\ref{eqn:tlocal})]
from the occupied orbital $|1  \rangle_1$
to the subspace of unoccupied orbitals $| 2 \rangle$ at the sites $2$ and $2'$.
Then, by assuming that all weights of $w^2(x)$ are accumulated at the lattice points
(that is the meaning of the ``lattice model''), one can write that
$$
w^2(x) = (1-q_- - q_+) \delta(x) + q_- \delta(x+a/2) + q_+ \delta(x-a/2),
$$
where
$$
q_{\pm} = \left(
\frac{\sin 2 \beta \mp | \sin \beta |}{2 \Delta}
\right)^2
$$
are the weights of $w^2(x)$ at the sites $2$ and $2'$.
By substituting this $w^2(x)$ into Eq.~(\ref{eqn:PWannier}),
we again arrive at Eq.~(\ref{eqn:Pinfinity}). Thus, in terms of these arguments,
the polarization is finite because $q_+ \ne q_-$.
Alternatively, one can say that due to the asymmetric electron transfer, the Wannier centers
are shifted from the centrosymmetric atomic positions.\cite{Yamauchi}
For a given $\Delta$, the difference
$(q_+ - q_-)$ depends on the value of $\beta$ and takes the maximal value
when $| \beta | = | \beta_{\rm max} | = \arctan \sqrt{2}$ (about $54.7^\circ$).

\begin{figure}
\begin{center}
\includegraphics[height=10cm]{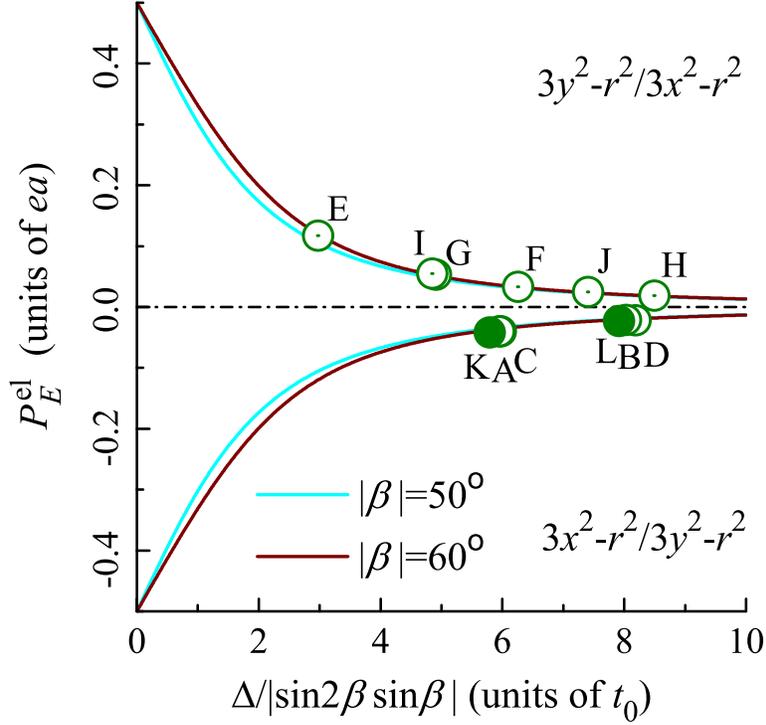}
\end{center}
\caption{\label{fig.PModel}(Color online)
Electronic polarization (more precisely -- the electric dipole moment) for the isolated zigzag chain as the function
of intra-atomic energy splitting between $e_g$ states. Upper part corresponds to the
$3y^2$$-$$r^2$/$3x^2$$-$$r^2$ type of the orbital ordering ($\beta$$>$$0$) and lower part
corresponds to the $3x^2$$-$$r^2$/$3y^2$$-$$r^2$ type of the orbital ordering ($\beta$$<$$0$).
The values obtained for YMnO$_3$ are shown by open symbols.
The points A, C, E, G, I denote the bare LDA
values, obtained for the experimental $P2_1nm$ and $Pbnm$ structures, and three theoretical
structures, obtained in LSDA and LDA$+$$U$ with $U$$=2.2$ and $6.0$ eV, respectively.
Similar values, obtained after adding the HF potentials,
are denoted as B, D, G, H, and J, respectively.
The values obtained for the experimental $Pbnm$ structure of HoMnO$_3$ are shown by filled symbols:
the point K denotes the bare LDA value and the point L takes into account the effect of the HF potential.}
\end{figure}

  The behavior of electronic polarization as the function of intra-atomic energy splitting
between $e_g$ states is summarized in Fig.~\ref{fig.PModel}.

  A very similar model of the FE polarization
in orthorhombic manganites was considered by Barone \textit{et al}.\cite{Barone}
The advantage of our approach is that we were able to reduce the problem to the
$2$$\times$$2$ Hamiltonian in the reciprocal space and to solve
it analytically. Such an analysis provides a transparent physical picture for the behavior of the
FE polarization. Therefore, we would like to stress briefly the difference
between our results and the ones by Barone \textit{et al}.
First, the behavior of polarization, obtained by Barone \textit{et al.}, is very
different from ours: it is zero for $\Delta = 0$ and approaches $\pm ea/2$ for
$\Delta \rightarrow \infty$. Nevertheless, such a difference can be easily understood by
the different choice of the reference point in the calculations of $P_E^{\rm el}$: $\Delta \rightarrow \infty$
in our work and $\Delta = 0$ in the work of Barone \textit{et al}. Another
discrepancy is related to the functional dependence of the orbital ordering and the FE polarization on $\Delta$:
in the work of Barone \textit{et al.},
these two quantities become finite starting only from some critical value of $\Delta$.
We believe that such a behavior is counterintuitive (at least, in the framework of the considered model) and
the orbital ordering, as well as the FE polarization, should evolve continuously starting from
$\Delta = 0$ (see also the analysis of the orbital ordering for similar model,
reported in Ref.~\onlinecite{Hotta}).

\subsection{\label{subsec:parameters} Parameters of the $e_g$ model and values of
electronic polarization for YMnO$_3$ and HoMnO$_3$}

  In this section, we evaluate parameters of the $e_g$ model for realistic compounds,
such as YMnO$_3$ and HoMnO$_3$. For these purposes, we do the following:
\begin{itemize}
\item[(i)]
We start with the realistic low-energy model,
derived for the Mn $3d$ bands of YMnO$_3$ and HoMnO$_3$ on the basis of
first-principles electronic structure calculations
(results of Refs.~\onlinecite{JPSJ} and \onlinecite{PRB12});
\item[(ii)]
Then, we
pick up parameters of the model for the single zigzag chain,
propagating along the orthorhombic $\boldsymbol{a}$ axis (and assuming that,
in the DE limit,
all transfer integrals in the directions $\boldsymbol{b}$ and $\boldsymbol{c}$
are blocked by the $E$-type AFM ordering);
\item[(iii)]
Solve the electronic structure problem for the isolated zigzag chain; find
eigenvalues and eigenfunctions;
\item[(iv)]
Construct the Wannier functions for the upper lying $e_g$ bands.
For these purposes, we use the projector-operator
technique and trial orbitals, obtained from the diagonalization of the site-diagonal part
of the density matrix;\cite{review2008}
\item[(v)]
Find parameters of the $e_g$ model in the obtained Wannier basis;
\item[(vi)]
Transform the parameters to the crystal-field representation, which diagonalizes the
site-diagonal part of the $e_g$ model;
\item[(vii)]
Fit the transfer integrals for the bond $1$-$2'$ in terms of $t_0$ and $\beta$, by using the
functional dependence given by Eq.~(\ref{eqn:tlocal}). Meanwhile, the splitting $\Delta$
between the $e_g$ states is obtained from the site-diagonal part.
\end{itemize}

  For YMnO$_3$, we have considered several crystal structures, which were previously
discussed in Ref.~\onlinecite{PRB12}:
\begin{itemize}
\item[(i)]
The experimental $Pbnm$ and $P2_1nm$ structures, reported in Ref.~\onlinecite{Okuyama};
\item[(ii)]
Three theoretical $P2_1nm$ structures, which were optimized in the local-spin-density approximation (LSDA) and
LDA$+$$U$ with $U$$=$ $2.2$ and $6.0$ eV by assuming the collinear
$E$-type AFM alignment without SO interaction. The results of this optimization can be found in Ref.~\onlinecite{PRB12}.
\end{itemize}
For HoMnO$_3$, we use the experimental $Pbnm$ structure, reported in Ref.~\onlinecite{Munoz}.

\begin{figure}
\begin{center}
\includegraphics[height=10cm]{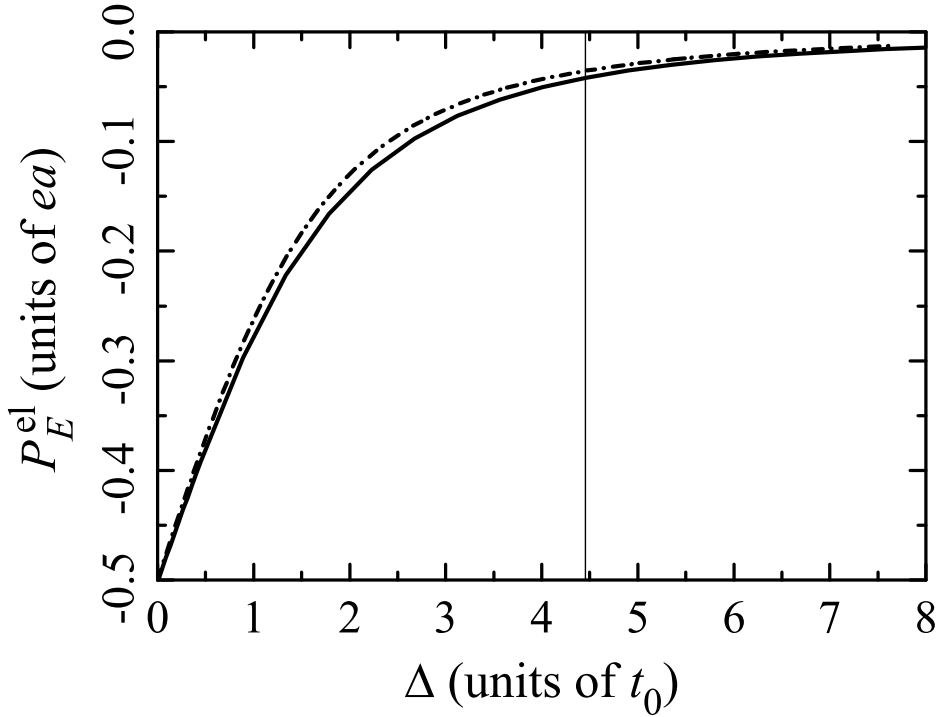}
\end{center}
\caption{\label{fig.compare}
Electronic polarization (more precisely -- the electric dipole moment) in the $e_g$ model for the isolated zigzag chain as
obtained by using bare LDA parameters for HoMnO$_3$ (solid line) and after
the parametrization of transfer integrals using Eq.~(\ref{eqn:tlocal}).
The vertical line shows the bare LDA value of $\Delta/t_0$.
The parameters are taken from Table~\ref{tab:zzparam}.}
\end{figure}

  Parameters of the $e_g$ model, obtained from the fitting, are summarized in Table~\ref{tab:zzparam}.
\begin{table}[h!]
\caption{Parameters of the $e_g$ model for the isolated zigzag chain, derived for
HoMnO$_3$ (HMO) and different
structures of YMnO$_3$ (YMO): the experimental $Pbnm$ and $P2_1nm$ structures,
reported in Ref.~\onlinecite{Okuyama}, and three $P2_1nm$ structures,
which were theoretically optimized
in LSDA and LDA$+$$U$ with $U$$=$ $2.2$ and $6.0$ eV by assuming the collinear
$E$-type AFM alignment (results of Ref.~\onlinecite{PRB12}).
In this Table, $t_0$ is the effective two-center integral $dd\sigma$,
$\Delta$ is the intra-atomic splitting between $e_g$ states, and $\beta$
specifies the form of the transfer integrals in the Mn-Mn bonds. The values, obtained
by using bare LDA parameters
are denoted as `LDA', and the ones after
adding the Hartree-Fock
potential are denoted as `$+$$U$'.}
\label{tab:zzparam}
\begin{ruledtabular}
\begin{tabular}{lcccccc}
 & \multicolumn{2}{c}{$t_0$ (meV)} & \multicolumn{2}{c}{$\Delta$ (eV)}
 & \multicolumn{2}{c}{$ \beta $ (degrees)} \\
 \cline{2-3} \cline{4-5} \cline{6-7}
                & LDA  & $+$$U$ & LDA & $+$$U$ & LDA  & $+$$U$ \\
\hline
HMO ($Pbnm$, Exp.)           & $341$ & $353$ & $1.52$ & $2.15$ & $-$$54.0$ & $-$$55.3$  \\
YMO ($Pbnm$, Exp.)           & $335$ & $348$ & $1.53$ & $2.18$ & $-$$54.2$ & $-$$55.8$  \\
YMO ($P2_1nm$, Exp.)         & $334$ & $346$ & $1.54$ & $2.15$ & $-$$54.1$ & $-$$55.7$  \\
YMO ($P2_1nm$, LSDA)         & $405$ & $412$ & $0.92$ & $1.95$ & $\phantom{-}$$57.6$ & $\phantom{-}$$59.1$  \\
YMO ($P2_1nm$, $U$$=2.2$ eV) & $361$ & $370$ & $1.37$ & $2.41$ & $\phantom{-}$$55.1$ & $\phantom{-}$$57.3$  \\
YMO ($P2_1nm$, $U$$=6.0$ eV) & $348$ & $359$ & $1.30$ & $2.04$ & $\phantom{-}$$54.4$ & $\phantom{-}$$56.1$  \\
\end{tabular}
\end{ruledtabular}
\end{table}
Eq.~(\ref{eqn:tlocal}) captures main details of transfer integrals between the nearest neighbors.
The largest deviation from the ideal $| \beta | = 60^\circ$ case was found
if one uses the bare LDA parameters,
derived for the experimental $Pbnm$ structure
of HoMnO$_3$. In this case, the agreement between the original matrices $\hat{\mathsf{t}}_{12'}$
and results of the fitting using Eq.~(\ref{eqn:tlocal})
is the worst:
$$
\hat{\mathsf{t}}_{12'} =
\left(
\begin{array}{rr}
119 & -328 \\
-12 & 39   \\
\end{array}\right) \qquad
\textrm{and} \qquad
\left(
\begin{array}{rr}
153 & -300 \\
-24 & 47   \\
\end{array}\right),
$$
before and after the fitting, respectively, in units of meV. On the other hand, $ \beta $
becomes close to $60^\circ$ if one uses theoretical LSDA crystal structure of YMnO$_3$ and takes into account the
additional level splitting, caused by the HF potential. In this case, the agreement between
the original and fitted matrices is nearly perfect. Nevertheless, we would like to emphasize that
the analytical expression, given by Eq.~(\ref{eqn:Pfinal}), with the parameters, derived from the fitting,
excellently reproduces the behavior of electronic polarization, obtained in the same $e_g$ model
but without fitting (see Fig.~\ref{fig.compare}). Thus, deviations of transfer integrals
from Eq.~(\ref{eqn:tlocal})
are
relatively unimportant for the
analysis of the FE polarization.

  It is interesting to note that $|P_E^{\rm el}| = ea/2$ when $\Delta \rightarrow 0^+$
(see Fig.~\ref{fig.compare}). This means that in the limit $\Delta \rightarrow 0$, the
system behaves such as if it would be centrosymmetric with respect to the
bond centers,\cite{Zak} even despite the fact that the space groups $Pbnm$ and
$P2_1nm$ (so as the transfer integrals) do not have such symmetry.
Apparently, such a behavior is related to a more general symmetry of the
transfer integrals.

  Then, we take the values of FE polarization, obtained for the $e_g$ band
(without fitting) and also plot them on Fig.~\ref{fig.PModel}. As
for the abscissa coordinates, we use results of Table~\ref{tab:zzparam}.
We can clearly see that all
these values fall on the analytical dependence, derived for the $e_g$ model.
The main parameter, which controls the value of the FE polarization,
is the ratio $\Delta/t_0$. The $\beta$-dependence of $P_E^{\rm el}$ is less important.
This result is very natural and will be discussed in a moment.
Moreover,
the physically relevant situation, realized in the
orthorhombic manganites, always corresponds to the
limit of large $\Delta$. This is another important finding, which will allow us to further
rationalize the behavior of the FE polarization in Sec.~\ref{subsec:5orbitals}.

  The polarization has different sign for the experimental
and theoretical structures, that indicates at different types of the orbital ordering
in the zigzag chain. In the $Pbnm$ phase, all zigzag chains are equivalent,
and in Fig.~\ref{fig.PModel} we simply picked up
the one with the same orbital ordering as in the $P2_1nm$ phase. However, in the
$P2_1nm$ phase, the type of the zigzag chain is uniquely defined (as the one with larger Mn-Mn
distances, which stabilize the FM coupling in the zigzag chain). Therefore, the sign difference between experimental
and theoretical values of $P_E^{\rm el}$ in the $P2_1nm$ phase
indicates at a serious problem, which may exist
in the first-principles calculations.
The problem will be discussed in details in Sec.~\ref{subsec:problems}.

\begin{figure}
\begin{center}
\includegraphics[height=10cm]{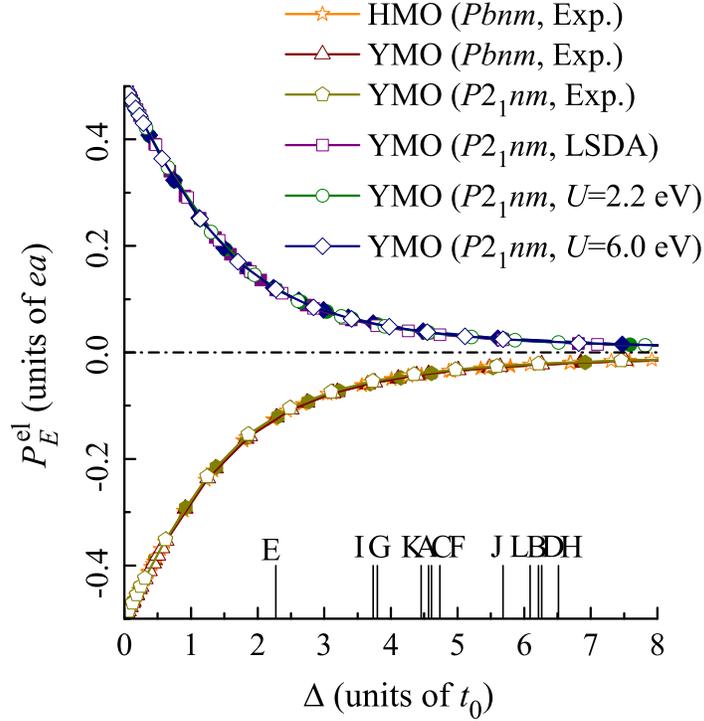}
\end{center}
\caption{\label{fig.PvsD}(Color online)
Electronic polarization (more precisely -- the electric dipole moment) versus $\Delta/t_0$, as obtained
using various sets of parameters for the $e_g$ model. Results of bare LDA
and after including the Hartree-Fock potential are shown by filled and
open symbols, respectively. The positions of $\Delta/t_0$ for different systems are shown by
capital letters. The points A, C, E, G, I stand for the bare LDA
values of $\Delta/t_0$, corresponding to the experimental $P2_1nm$ and $Pbnm$ structures, and three theoretical
structures, obtained in LSDA and LDA$+$$U$ with $U$$=2.2$ and $6.0$ eV, respectively.
Similar points, obtained after adding the Hartree-Fock potential,
are denoted as B, D, G, H, and J, respectively. The points K and L correspond to the
$Pbnm$ structure of HoMnO$_3$, obtained in the bare LDA and after including the HF potential, respectively.}
\end{figure}

  Then, all values of $| \beta |$ are close to $|\beta_{\rm max}| \approx 54.7^\circ$,
which corresponds to the maximum of $| P_E^{\rm el} |$ (see Table~\ref{tab:zzparam}).
Therefore, any deviation of $P_E^{\rm el}(\beta)$ from $P_E^{\rm el}(\beta_{\rm max})$ will be only
of the order of $(\beta$$-$$\beta_{\rm max})^2$.
Thus, all the effects of $\beta$ on $P_E^{\rm el}$ will be small.
This can be clearly seen in Fig.~\ref{fig.PvsD},
where we plot $ P_E^{\rm el} $ versus $\Delta/t_0$, using different
sets of parameters for the $e_g$ model and varying $\Delta$:
all lines, corresponding to different crystal structures and different levels of approximation for the
on-site interactions (with and without the HF potential), are practically
undistinguishable. This means that, in reality, $P_E^{\rm el}$ is controlled by only two sets of parameters:
(i) the ratio $\Delta/t_0$, and (ii) the lattice parameters $a$, $b$, and $c$,
which determine the value of the scaling factor $a/V$ in the polarization density.
The $\beta$-dependence of $P_E^{\rm el}$ is relatively unimportant.

  From the physical point of view, the $ \beta $-dependence
of the transfer integrals is related to the buckling of the Mn-O-Mn bonds. Then, the above result
suggests that $P_E^{\rm el}$ does not explicitly depend on the Mn-O-Mn angles: the latter can contribute
to $P_E^{\rm el}$, but only via
other model parameters (such as $t_0$), which depend on these angles. This finding is
consistent with the conclusion of Ref.~\onlinecite{Yamauchi}, based on the
first-principles electronic structure calculations.

  Finally, we briefly explain the correspondence between the values of the
electric dipole moment in Fig.~\ref{fig.PModel} and the polarization density. Let us consider the
experimental $Pbnm$ structure of YMnO$_3$. Then, the value $-$$0.022ea$,
which takes into account the effect of the HF potential, corresponds to the polarization
density of about $-1.65$ $\mu$C/cm$^2$. It should be remembered that it is only the contribution
of the $e_g$ band alone. In order to obtain the total polarization for the five-orbital model,
it should be combined with the contribution of the $t_{2g}$ band. This yields
the total polarization $-0.84$ $\mu$C/cm$^2$,
which agrees with the value for the $E$-type AFM state (for $\phi = 180^\circ$) in Fig.~\ref{fig.Ecanting}.
Thus, the contributions of the $t_{2g}$ and $e_g$ bands have opposite sign and partially
cancel each other, in agreement with the first-principles calculations.\cite{Yamauchi}
In the rest of this work, we will deal with the total polarization density, including the
effect of both $t_{2g}$ and $e_g$ bands.

\subsection{\label{subsec:5orbitals} Electronic polarization in the five-orbital
model: simple analytical expression}

  Now, we will generalize results of two previous sections and derive an approximate, but very
transparent expression for the electronic polarization in orthorhombic manganites with a general twofold periodic
magnetic texture. Our starting point is that the behavior of electronic
polarization in realistic compounds corresponds to the limit of large $\Delta$.
This limit can be justified even without on-site Coulomb interactions
(i.e., considering the ratio of transfer integrals to the crystal-field splitting in bare LDA),
and is additionally strengthened after including the Coulomb interactions. Thus, the central
quantity, which we should evaluate in the second order of $1/\Delta$,
is the weight $w_{i \rightarrow j}^2$, transferred from the Wannier
orbital at the site $i$ to the neighboring site $j$.
Moreover,
since electronic polarization is equal to zero for the fully occupied band, it is more
convenient to start with the unoccupied $e_g$ orbitals
and consider the transfer integrals to the subspace of
three $t_{2g}$ and one $e_g$ occupied orbitals at each of the neighboring sites.
This procedure should give us $-$${\bf P}^{\rm el}$.

\begin{figure}
\begin{center}
\includegraphics[height=5cm]{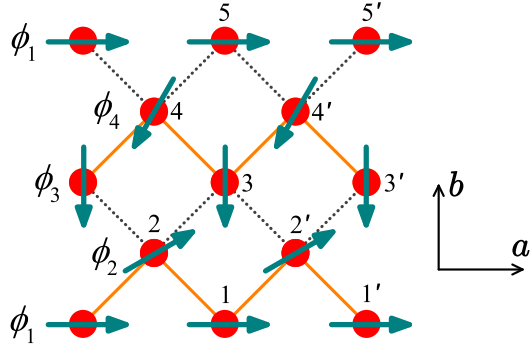}
\end{center}
\caption{\label{fig.2Dtexture}(Color online)
General twofold periodic magnetic texture in the $\boldsymbol{ab}$ plane
of orthorhombic manganites, which
remains invariant under the symmetry
operation $\hat{S} = \{ \hat{C}^2_a| \boldsymbol{a}/2$$+$$\boldsymbol{b}/2 \}$
of the space group $Pbnm$.
Solid and dotted lines denote two types of magnetically inequivalent bonds.}
\end{figure}

  The transfer integrals obey certain symmetry rules and, in the DE model,
are additionally modulated by $\xi_{ij}$. More specifically, we consider a planar magnetic texture
which is shown in Fig.~\ref{fig.2Dtexture}.
The periodicity of this texture along the orthorhombic axes is $a$ and $2b$, respectively.
The directions of spins are specified by three azimuthal angles: $\phi_2$, $\phi_3$, and $\phi_4$
(while $\phi_1 = 0$ is treated as the reference point). Moreover, we assume that the DE Hamiltonian
remains invariant under the symmetry operation
$\hat{S} = \{ \hat{C}^2_a| \boldsymbol{a}/2$$+$$\boldsymbol{b}/2 \}$,
which transforms the bond $1$-$2$ to $4'$-$3$, the bond
$3$-$2$ to $4$-$5$, etc.
In the DE model,
it imposes additional conditions on the azimuthal angles:
$\cos \frac{\phi_2}{2} = \cos \frac{\phi_3 - \phi_4}{2}$ and $\cos \frac{\phi_4}{2} = \cos \frac{\phi_3 - \phi_2}{2}$,
which are satisfied if $\phi_3 = \phi_2$$\pm$$\phi_4$ (modulo $2 \pi$).
Thus, the magnetic texture is specified by only two independent parameters $\phi_2$ and $\phi_4$,
similar to the magnetic texture obtained in the mean-field HF calculations with the SO coupling.\cite{PRB11,PRB12}

\begin{figure}
\begin{center}
\includegraphics[height=8cm]{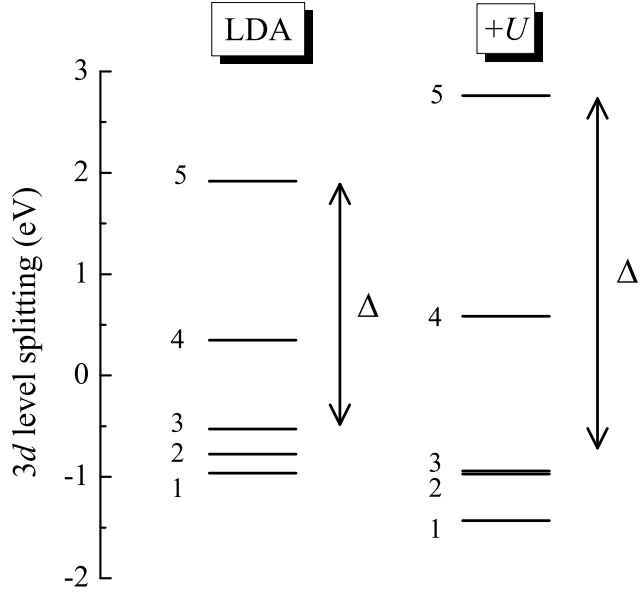}
\end{center}
\caption{\label{fig.CF}
Splitting of $3d$ levels for the experimental $Pbnm$ phase of YMnO$_3$.
The values, obtained
using bare LDA parameters of the low-energy model
are denoted as `LDA', and the ones obtained after
adding the Hartree-Fock
potential are denoted as `$+$$U$'. $\Delta$ is the energy splitting between the unoccupied orbital $5$
and the center of gravity of occupied orbitals $1$-$4$.}
\end{figure}

  Then, we consider some central site (say, site $3$ in Fig.~\ref{fig.2Dtexture}) and evaluate its
contribution to the electronic polarization, which is caused by the Wannier weight transfer to the neighboring sites
$4'$, $4$, $2$, and $2'$, located at $(\boldsymbol{a}$$+$$\boldsymbol{b})/2$, $-$$(\boldsymbol{a}$$-$$\boldsymbol{b})/2$,
$-$$(\boldsymbol{a}$$+$$\boldsymbol{b})/2$, and $-$$(\boldsymbol{a}$$-$$\boldsymbol{b})/2$, respectively.
In the second order of $1/\Delta$
(and apart from the proportionality coefficient, which will be specified later),
the contribution of the site $3$ to the vector of electronic polarization can be written as
{\setlength\arraycolsep{2pt}
\begin{eqnarray}
{\bf P}_3^{\rm el} & \sim & \frac{e}{2} \cos^2 \frac{\phi_2}{2}
\left[
( \boldsymbol{a} + \boldsymbol{b} ) w_{3 \rightarrow 4'}^2 -
( \boldsymbol{a} - \boldsymbol{b} ) w_{3 \rightarrow 4}^2
\right]
+ {} \nonumber \\
 & & {}
\frac{e}{2} \cos^2 \frac{\phi_4}{2}
\left[
( \boldsymbol{a} - \boldsymbol{b} ) w_{3 \rightarrow 2'}^2 -
( \boldsymbol{a} + \boldsymbol{b} ) w_{3 \rightarrow 2}^2
\right],
\label{eqn:P31}
\end{eqnarray}}
where
$w_{i \rightarrow j}^2$ is proportional to the sum of squares of the transfer integrals from the unoccupied orbital 5 at the site $i$ to the
occupied orbitals $1$-$4$ at the site $j$:
$w_{i \rightarrow j}^2 = \left[ ( \mathsf{t}_{ij}^{51} )^2 + ( \mathsf{t}_{ij}^{52} )^2
 + ( \mathsf{t}_{ij}^{53} )^2 + ( \mathsf{t}_{ij}^{54} )^2 \right]/\Delta^2$.
These transfer integrals should be calculated in the
`crystal-field representation', that diagonalizes the site-diagonal part of the one-electron
Hamiltonian. The parameter $\Delta$ is understood as the energy difference between the unoccupied
orbital $5$ and the center of gravity of occupied orbitals 1-4 (see Fig.~\ref{fig.CF}).
Thus, in this analysis, we neglect
the splitting between the occupied orbitals, which is smaller than $\Delta$.
Then, in the $Pbnm$ structure, each Mn site is located in the inversion center. Therefore,
$w_{i \rightarrow j}^2$ in the bonds
$3$-$4'$ and $3$-$2$ (as well as $3$-$2'$ and $3$-$4$) are equivalent, and Eq.~(\ref{eqn:P31}) can be further transformed to
\begin{equation}
{\bf P}_3^{\rm el} \sim \frac{e}{4} \left( \cos \phi_2 - \cos \phi_4 \right)
\left[
( \boldsymbol{a} + \boldsymbol{b} ) w_{3 \rightarrow 4'}^2 -
( \boldsymbol{a} - \boldsymbol{b} ) w_{3 \rightarrow 4 }^2
\right].
\label{eqn:P32}
\end{equation}
Similar analysis can be performed for another Mn site in the primitive cell (say, site $4'$ in Fig.~\ref{fig.2Dtexture}).
Moreover, since the sites $3$ and $4'$ are connected by the symmetry operation
$\hat{S} = \{ \hat{C}^2_a| \boldsymbol{a}/2$$+$$\boldsymbol{b}/2 \}$, using Eq.~(\ref{eqn:P32}), one can immediately
obtain that
$$
{\bf P}_{4'}^{\rm el} \sim \frac{e}{4} \left( \cos \phi_2 - \cos \phi_4 \right)
\left[
( \boldsymbol{a} - \boldsymbol{b} ) w_{3 \rightarrow 4'}^2 -
( \boldsymbol{a} + \boldsymbol{b} ) w_{3 \rightarrow 4}^2
\right].
$$
Then, the total polarization ${\bf P}^{\rm el} = 2({\bf P}_3^{\rm el}$$+$${\bf P}_{4'}^{\rm el})$
can be evaluated as
$$
{\bf P}^{\rm el} = \frac{e}{V} \left( \cos \phi_2 - \cos \phi_4 \right)
\left[
 w_{3 \rightarrow 4'}^2 -
 w_{3 \rightarrow 4}^2
\right] \boldsymbol{a}.
$$
Here, $V$ is the primitive cell volume, containing four Mn sites
(two in each of the $\boldsymbol{ab}$ planes, located at $z$$=$ $0$ and $c/2$, which is reflected in the
additional prefactor $2$ in the above expression). Finally, by applying the symmetry operation
$\hat{S} = \{ \hat{C}^2_a| \boldsymbol{a}/2$$+$$\boldsymbol{b}/2 \}$, the sites $3$ and $4$ can be
transformed to the sites $4'$ and $3$, respectively. Thus, ${\bf P}^{\rm el}$ can be expressed through the transfer
integrals in only one nearest-neighbor (NN) bond $3$-$4'$ (or in any equivalent to it bond):
\begin{equation}
{\bf P}^{\rm el} = \frac{1}{2}\left( \cos \phi_2 - \cos \phi_4 \right)
{\bf P}_E^{\rm el},
\label{eqn:Ptotal}
\end{equation}
where
\begin{equation}
{\bf P}_E^{\rm el} = \frac{2e}{V}
\left[
 w_{3 \rightarrow 4'}^2 -
 w_{4' \rightarrow 3}^2
\right] \boldsymbol{a}
\label{eqn:PE}
\end{equation}
is the electronic polarization in the $E$-type AFM state.
For an arbitrary direction of spin at the site $1$, the angular dependence
$\left( \cos \phi_2 - \cos \phi_4 \right)$ in Eq.~(\ref{eqn:Ptotal}) should be replaced by a more general expressions
${\bf e}_1 \cdot \left( {\bf e}_2 - {\bf e}_4 \right)$.
Eqs.~(\ref{eqn:Ptotal}) and (\ref{eqn:PE}) allow us to rationalize many aspects of the multiferroic activity in manganites
with the twofold periodic magnetic texture, namely:
\begin{itemize}
\item[(i)]
${\bf P}^{\rm el}$ is parallel to the orthorhombic $\boldsymbol{a}$ axis;
\item[(ii)]
If $\phi_4 = \phi_2$$+$$\pi$, ${\bf P}^{\rm el}$ is proportional to $\cos \phi_2$, which nicely explains the functional
dependence of ${\bf P}^{\rm el}(\phi)$ in Fig.~\ref{fig.Ecanting}(b) and
in the first-principles calculations for the same magnetic geometry (Ref.~\onlinecite{Picozzi});
\item[(iii)]
${\bf P}^{\rm el}$ vanishes in
the homogeneous spin-spiral state ($\phi_2 = \pi/2$ and $\phi_4 = 3 \pi/2$).
This is a very natural result from the viewpoint of the
DE physics: in the spin-spiral texture, all $|\xi_{ij}|$ are the same. Therefore,
all bonds remain equivalent, and the inversion symmetry of the DE Hamiltonian is not broken;
\item[(iv)]
Since $\hat{\mathsf{t}}_{ji} = \hat{\mathsf{t}}_{ij}^T$, ${\bf P}_E^{\rm el}$ can be also presented
in the form
\begin{equation}
{\bf P}_E^{\rm el} = \frac{2e}{V}
\frac{ \left( \vec{v}_+ , \vec{v}_- \right)}{\Delta^2}
\boldsymbol{a},
\label{eqn:PtotalE}
\end{equation}
where $\left( \vec{v}_+ , \vec{v}_- \right)$
is the scalar product of the 4-dimensional vectors
$\vec{v}_{\pm} \equiv (v^1_{\pm},v^2_{\pm},v^3_{\pm},v^4_{\pm})$, constructed from symmetric ($+$) and antisymmetric ($-$)
parts of the transfer integrals:
$v_{\pm}^m = \mathsf{t}_{ij}^{5m}$$\pm$$\mathsf{t}_{ij}^{m5}$.
Thus, in order to have finite ${\bf P}^{\rm el}$, the matrix of transfer integrals
should
have both symmetric
and antisymmetric components.
\end{itemize}

  Let us evaluate ${\bf P}_E^{\rm el} \equiv (P_E^{\rm el},0,0)$, using Eq.~(\ref{eqn:PtotalE}), for the
experimental $Pbnm$ phase of YMnO$_3$. In this case, the unit cell volume is
$V = 224.13$ \AA$^3$ and the orthorhombic lattice parameter is $a = 5.245$ \AA.\cite{Okuyama} Then, for the
bare LDA band structure, we have: $\Delta = 2.40$ eV (see Fig.~\ref{fig.CF}), $\vec{v}_+ = (-$$125, 18, 15, 336)$ meV,
and $\vec{v}_- = (99,-$$49,-$$25,-$$314)$ meV (all parameters of the low-energy model for YMnO$_3$
can be found in Supplemental Material of Ref.~\onlinecite{PRB12}).
By substituting all these values in Eq.~(\ref{eqn:PtotalE}), we obtain $P_E^{\rm el} = -$$1.55$ $\mu$C/cm$^2$, which
agrees very well with the value of $-$$1.53$ $\mu$C/cm$^2$, obtained directly from the Berry-phase formula
[Eq.~(\ref{eqn:PKSV})], without
additional approximations (apart from the DE limit). For the more realistic case, including the
effect of the HF potential, we have: $\Delta = 3.45$ eV, $\vec{v}_+ = (6,-$$117,26,335)$ meV,
and $\vec{v}_- = (10,91,-$$24,-$$319)$ meV. Then, Eq.~(\ref{eqn:PtotalE}) yields
$P_E^{\rm el} = -$$0.74$ $\mu$C/cm$^2$, which is again consistent with the value of $-$$0.85$ $\mu$C/cm$^2$,
obtained directly from the Berry phase formula [Eq.~(\ref{eqn:PKSV})].
Moreover, the values of the scalar product $\left( \vec{v}_+ , \vec{v}_- \right)$
appear to be very close when they are calculated
with and without the HF potential: $-$$0.118$ and $-$$0.119$ eV$^2$, respectively.
This result is very natural because the form of the crystal-field orbitals in orthorhombic manganites
is mainly controlled by the JT distortion: the latter is large and thus `decides' which orbitals will be occupied and
which will not. On the other hand, the effect of on-site Coulomb interactions, being inversely proportional to $U$,\cite{KugelKhomskii}
is considerably smaller. Thus, although the Coulomb interactions contribute to the splitting between
occupied and empty states (see Fig.~\ref{fig.CF}), they
practically do not change the subspace of occupied orbitals.
Therefore, the construction $\left( \vec{v}_+ , \vec{v}_- \right)$, which is evaluated in the crystal-field
representation, will not strongly depend on whether it is calculated with or without the HF potential.
In such a situation, the absolute value of $P_E^{\rm el}$ will be mainly controlled
by the parameter $\Delta$ in the denominator of Eq.~(\ref{eqn:PtotalE}).

  Furthermore, $\Delta$ can be presented in the form: $\Delta = \Delta_{\rm JT} + \Delta_U$, where $\Delta_{\rm JT}$
and $\Delta_U$ take into account the effects of the bare JT distortion and
the on-site Coulomb interactions, respectively. In the
example considered above, $\Delta_{\rm JT}$ is the LDA level splitting and $\Delta_U$ is
the additional splitting, caused by the HF potential (see Fig.~\ref{fig.CF}).
Then, if $P_E^{\rm el}(0)$ is the
FE polarization in LDA,
the effect of on-site Coulomb interactions on $P_E^{\rm el}$
can be evaluated using the following scaling relation:
$$
P_E^{\rm el}(\Delta_U) = P_E^{\rm el}(0)/(1+\Delta_U/\Delta_{\rm JT})^2,
$$
which was observed in many LDA$+$$U$ calculations, treating the on-site Coulomb repulsion $U$
as an adjustable parameter.\cite{Picozzi,Okuyama}

  Finally, it is instructive to evaluate ${\bf P}^{\rm el} \equiv (P^{\rm el},0,0)$
for the noncollinear magnetic ground state of YMnO$_3$
using Eq.~(\ref{eqn:Ptotal}). This magnetic ground state was obtained in Ref.~\onlinecite{PRB12} by
solving mean-field HF equations with the relativistic SO interaction. For the $Pbnm$ phase of YMnO$_3$,
it yields $\phi_2 = 60^\circ$ and $\phi_4 = 240^\circ$.
Then, using the value $P_E^{\rm el} = -$$0.85$ $\mu$C/cm$^2$, obtained in the DE limit (see Fig.~\ref{fig.Ecanting}),
$P^{\rm el}$ can be estimated as $-$$0.43$ $\mu$C/cm$^2$, which is consistent reasonably well with
$P^{\rm el} = -$$0.55$ $\mu$C/cm$^2$, obtained for the noncollinear magnetic
ground state of YMnO$_3$ without additional approximations.\cite{PRB12}
In fact, the main discrepancy is caused by the DE limit for $P_E^{\rm el}$. For example,
if one uses $P_E^{\rm el} = -$$1.04$ $\mu$C/cm$^2$, obtained without
the DE approximation,\cite{PRB12} and the angular dependence
of ${\bf P}^{\rm el}$, given by Eq.~(\ref{eqn:Ptotal}), $P^{\rm el}$ can be estimated as $-$$0.50$ $\mu$C/cm$^2$,
which is much closer to $P^{\rm el} = -$$0.55$ $\mu$C/cm$^2$.

\subsection{\label{subsec:problems} Relative directions of electronic and ionic polarization,
and problems of structural optimization in
LDA$+$$U$}

  So far, we considered only electronic polarization, which was induced
by the orbital ordering in the FM zigzag chains. In this section, we will discuss how this
electronic
part is related to the ionic polarization in the noncentrosymmetric $P2_1nm$ structure.

  Moreover, we will elucidate the microscopic origin of the ``order of magnitude difference'',
which typically exists between experimental and theoretical values of the FE polarization, reported
for the orthorhombic manganites with twofold periodic magnetic texture.
The problem is formulated as follows. The great advantage of
the first-principles calculations is that they allow us to perform the structural optimization and
to find theoretically the atomic displacements, which are caused by the exchange-striction
effects in the $E$-type AFM phase. If one does
such structural optimization for the orthorhombic manganites
and subsequently calculates the FE polarization,
the latter will be of the order of several $\mu$C/cm$^2$.\cite{Picozzi}
The conclusion is rather generic and was obtained for several popular types of the
exchange-correlation functionals, such as LSDA (Ref.~\onlinecite{PRB12}),
generalized gradient approximation (GGA, Refs.~\onlinecite{Picozzi,Yamauchi}),
and LDA(GGA)$+$$U$ (Refs.~\onlinecite{Picozzi,PRB12}). The experimental polarization is
typically smaller than $0.5$ $\mu$C/cm$^2$.\cite{Ishiwata} On the other hand, if one takes the
experimental $P2_1nm$ structure and calculates the FE polarization, it will be at least of the same
order of magnitude as the experimental one.\cite{Okuyama,PRB12} The reason
of such discrepancy is that,
in the experimental
$P2_1nm$ structure,
there is a large
cancelation of electronic and ionic contributions to the FE polarization,
while in the theoretically optimized structure, these two contributions have
the same sign and the cancelation does not occur.\cite{PRB12}

  In this section, we will further clarify the situation.
In orthorhombic manganites, there are three types of atomic displacements, which control the FE polarization:
\begin{itemize}
\item[(i)] The Jahn-Teller distortion, which
gives rise to the orbital ordering;
\item[(ii)] The exchange striction, which specifies the type of the ordering in the FM zigzag chain and,
therefore, the sign of the electronic polarization. Note, that in the cenrosymmetric $Pbnm$ structure,
the FM chains with the $3x^2$$-$$r^2$/$3y^2$$-$$r^2$ and $3y^2$$-$$r^2$/$3x^2$$-$$r^2$ type of the
orbital ordering are equivalent as they build two degenerate magnetic states.
This degeneracy is lifted in the $P2_1nm$ phase
by the exchange striction effects, which pick up only one type
of the FM zigzag chains (characterized by larger Mn-Mn distances).
As soon as the FM chains are selected, the type of the orbital ordering is fixed,
so as the sign of the electronic polarization.
\item[(iii)] The FE atomic displacements, which occur in response to the
magnetic inversion symmetry breaking and control the sign of the ionic polarization.
\end{itemize}
The goal of this section is to understand how these three types of the lattice distortions correlate with each other
in the experimental and theoretically optimized $P2_1nm$ structures of YMnO$_3$.

  Let us consider the ionic polarization and concentrate on the behavior of the oxygen sites,
which are located in the $\boldsymbol{ab}$ plane and give the
largest contribution to ${\bf P}_E^{\rm ion}$.\cite{PRB12}
In principles, one can consider the contributions of other atomic sites,
which do not alter the conclusions.
Then, ${\bf P}_E^{\rm ion}$ can be presented in the following form:
\begin{equation}
{\bf P}_E^{\rm ion} = \frac{1}{2V} \sum_i Z_i \Delta \boldsymbol{\tau}_i,
\label{eqn:ionicP}
\end{equation}
where $Z_i$ are the atomic charges and $\Delta \boldsymbol{\tau}_i$ are the
atomic displacements away from the centrosymmetric positions.
Moreover, it is understood that around each Mn site in the primitive cell,
the summation runs over four oxygen sites, located in the nearest neighborhood of Mn.
Since each oxygen is shared by two Mn atoms, this leads to the additional prefactor $1/2$.
There are many possibilities for choosing the centrosymmetric reference point for
evaluation of
$\Delta \boldsymbol{\tau}_i$. The final result should not depend on this choice.
For our purposes, it is convenient to choose
$\Delta \boldsymbol{\tau}_i = \boldsymbol{\tau}_{\rm O}$$-$$\boldsymbol{\tau}_{\rm Mn}$
(in the other words, we assume that in the centrosymmetric structure, all oxygen sites
``fall'' on the central Mn site). This can be done because Mn sites do not contribute to the
FE polarization of the ionic type along the orthorhombic $\boldsymbol{a}$ axis.\cite{PRB12}
The reason is that,
apart from a constant shift, the projections of Mn sites onto the $\boldsymbol{a}$ axis
are either $0$ or $a/2$ (modulo the lattice translation $a$) and, therefore,
can be transformed to each other by the reflection $a \rightarrow -$$a$.
The Mn sites do contribute to the ionic polarization in the $\boldsymbol{bc}$ plane. However, all
these contributions have antiferroelectric character and cancel out after summation over
the primitive cell. Thus, around each Mn site, the evaluation of ${\bf P}_E^{\rm ion}$
is reduced to the summation of $\Delta \boldsymbol{\tau}_i$ over neighboring Mn-O bonds with
the perfactors given by Eq.~(\ref{eqn:ionicP}).
Such a construction is very convenient, because in the centrosymmetric $Pbnm$ structure, each Mn site
is located in the inversion center. Therefore, the sum of $\Delta \boldsymbol{\tau}_i$
over all neighboring Mn-O bonds will be equal to zero. In the $P2_1nm$ structure, however, such a construction
will give us a finite vector, which can serve as a measure of noncentrosymmetric atomic
displacements around each Mn site.
For our purposes, only the FE ($\boldsymbol{a}$) components of these vectors are important, while
the $\boldsymbol{b}$ and $\boldsymbol{c}$ components are antiferroelectric and will cancel each other.
Using this construction and taking the ionic value $Z_{\rm O} = -$$2|e|$, the contribution of the
planar oxygen sites to $P_E^{\rm ion}$ in the experimental $P2_1nm$ structure can be
estimated as $0.73$ $\mu$C/cm$^2$, which is totally consistent with the previous finding.\cite{PRB12}

\begin{figure}
\begin{center}
\includegraphics[height=5cm]{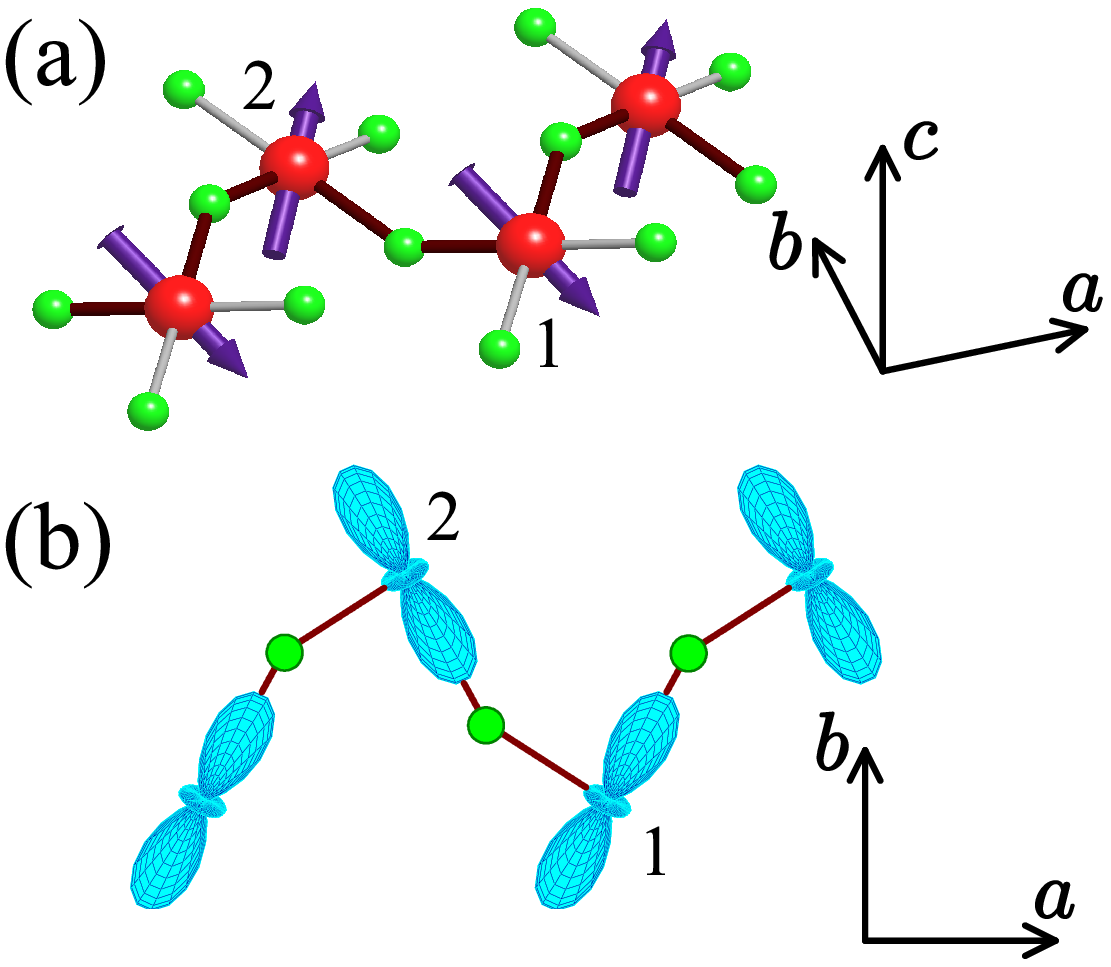}
\includegraphics[height=5cm]{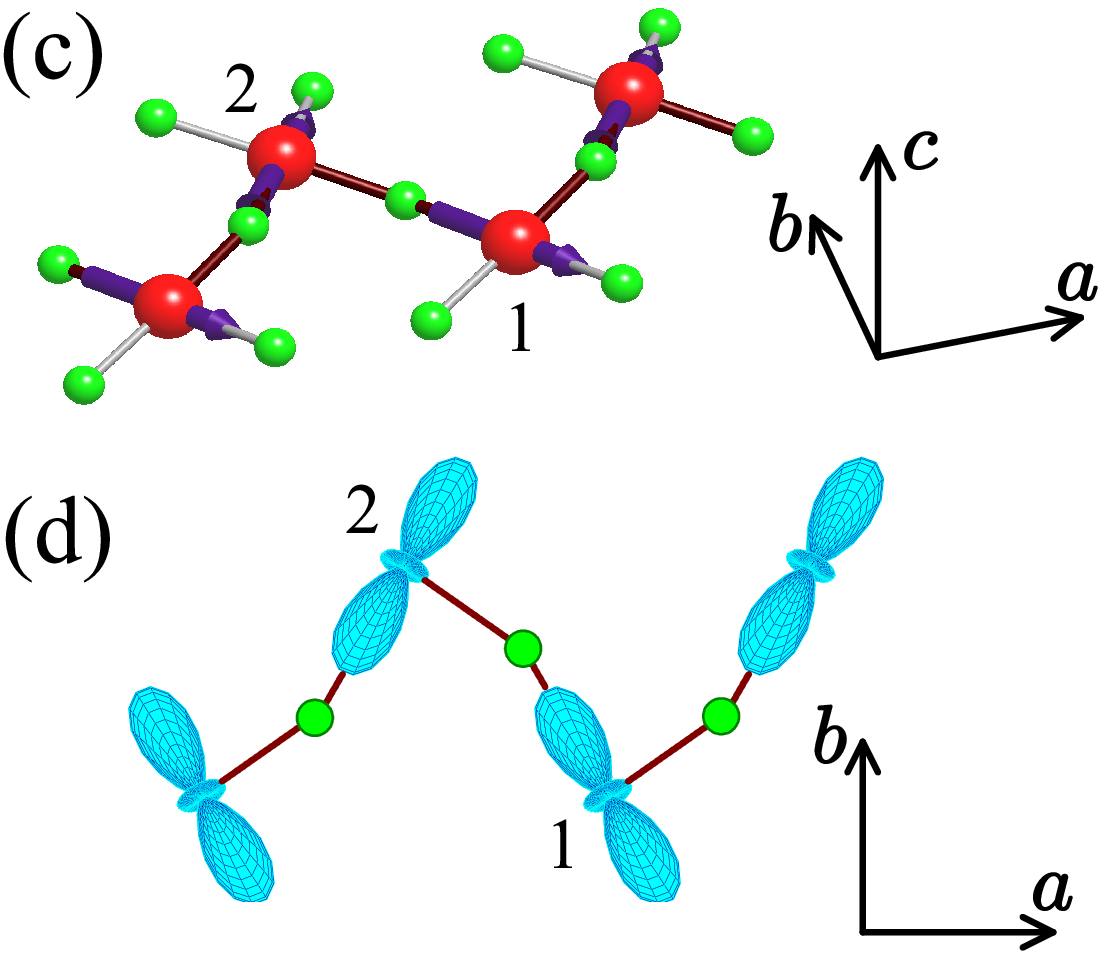}
\end{center}
\caption{\label{fig.problems} (Color online)
Directions of ionic contributions to the polarization, caused by ferroelectric displacements of
oxygen atoms around each Mn site in the $\boldsymbol{ab}$ plane of
noncentrosymmetric $P2_1nm$ phase of YMnO$_3$ (a and c), and the orbital
ordering, realized in the ferromagnetic zigzag chain (b and d), as obtained for the experimental (a and b)
and theoretically optimized structure (c and d).}
\end{figure}

  The distributions of such vectors, obtained for the experimental and theoretical structures
of YMnO$_3$,
are shown in Figs.~\ref{fig.problems}(a) and \ref{fig.problems}(c), respectively.
As for the theoretically optimized structure, we use results of LDA$+$$U$
calculations with $U$$=$ $2.2$ eV (see Ref.~\onlinecite{PRB12}).
Nevertheless, we would like to emphasize that very similar results were
obtained in LSDA and LDA$+$$U$ with $U$$=$ $6.0$ eV.\cite{PRB12}
As is seen in Fig.~\ref{fig.problems}, the FE displacements have the same direction in the
experimental and theoretically optimized $P2_1nm$ structure of YMnO$_3$.
This direction corresponds to the positive value of $P_E^{\rm ion}$.

  Corresponding orbital ordering, realized in the FM chains, is shown
in Figs.~\ref{fig.problems}(b) and \ref{fig.problems}(d),
for the experimental and theoretical structure, respectively.
For the experimental $P2_1nm$ structure,
the orbital ordering is of the $3x^2$$-$$r^2$/$3y^2$$-$$r^2$ type.
Therefore, the electronic polarization is negative, and there is a partial cancelation of the electronic and ionic terms,
which explains a relatively small value of the experimental polarization.\cite{PRB12}
However, the theoretical optimization of the crystal structures, performed both in
LSDA and LDA$+$$U$, yields different type of the orbital ordering:
$3y^2$$-$$r^2$/$3x^2$$-$$r^2$ instead of
$3x^2$$-$$r^2$/$3y^2$$-$$r^2$. Therefore, the electronic polarization will be positive, and the cancelation does not occur.

  Thus, the directions of FE displacements, obtained
in LSDA and LDA$+$$U$, are
inconsistent with the type of the orbital ordering, realized in the FM zigzag chains.
This seems to be a serious problem of the first-principles calculations and at the present stage it
is not clear how it should be solved. On the computational side, many attention recently is paid
to the screened hybrid functionals (see, e.g., Ref.~\onlinecite{HeFranchini}).
Therefore, it would be interesting to see how these functionals will work for the structural optimization
in multiferroic compounds, where the inversion symmetry is broken by the magnetic degrees of freedom.
The first applications for HoMnO$_3$ seem to show that the problem persists: although the
electronic polarization is decreased, mainly due to the increase of the
on-site level splitting, it has the same
sign as the ionic one and the total polarization is overestimated in comparison with the
experiment.\cite{Stroppa}
On the other hand, the directions of FE displacements can be controlled by the relativistic
SO interaction, which is typically ignored in the process of structural optimization.
This point of view was proposed, for example, in Ref.~\onlinecite{Malashevich}.

\subsection{\label{subsec:switching} Switching electric polarization
by changing the magnetic texture}

  What is interesting about the multiferroic systems is that the value and the direction
of the FE polarization depend on the magnetic texture and,
by changing this texture, one can also change the vector of polarization.
In this section, we will discuss how such a behavior can be realized in the twofold periodic magnetic texture.
Again, let us consider the centrosymmetric $Pbnm$ structure and assume that
the inversion symmetry is broken exclusively by the magnetic order.
In such a case,
most of attention is focused on the $E$-type AFM phase (Fig.~\ref{fig.Ecanting}),
which breaks
the inversion symmetry but preserves the symmetry operation
$\{ \hat{C}^2_a| \boldsymbol{a}/2$$+$$\boldsymbol{b}/2 \}$.
Therefore, the FE polarization will be parallel to the $\boldsymbol{a}$ axis.

\begin{figure}
\begin{center}
\includegraphics[height=10cm]{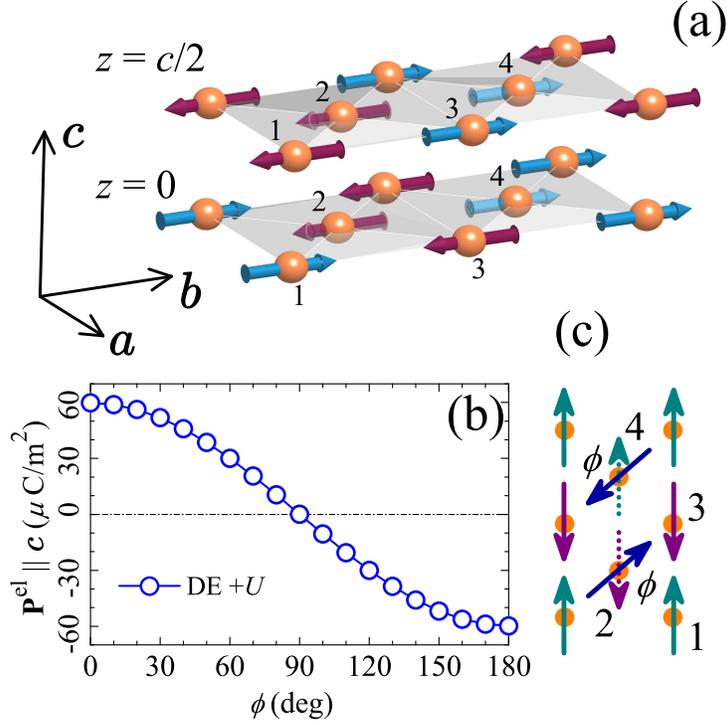}
\end{center}
\caption{\label{fig.iEcanting}(Color online)
(a) Antiferromagnetic texture yielding finite ferroelectric polarization
along the $\boldsymbol{c}$ axis. (b) Behavior of electronic polarization
in YMnO$_3$ upon
rotation of magnetic moments, as obtained in the
double exchange model with the Hartree-Fock potential $\hat{\cal V}_i^\uparrow$
(DE $+$$U$). In the rotated texture, the directions of spins
at the sites $1$ and $3$ were fixed, while the spins at the sites $2$ and $4$ were
rotated by the angle $\phi$, as explained in panel (c). The interlayer coupling was kept
AFM for the sites $1$ and $3$ and FM for the sites $2$ and $4$.
}
\end{figure}

  Now, the question is whether there are other
types of the magnetic texture, which would break the inversion symmetry.
As an example, let us consider the magnetic texture in Fig.~\ref{fig.iEcanting}(a).
In the plane $z=0$, it is identical to the $E$-type AFM order, and can be transformed to itself
by applying the symmetry operation $\{ \hat{C}^2_a| \boldsymbol{a}/2$$+$$\boldsymbol{b}/2 \}$ around
even magnetic sites $2$ and $4$. Alternatively, one can apply the symmetry operation
$\{ \hat{C}^2_a| -$$\boldsymbol{a}/2$$-$$\boldsymbol{b}/2 \}$ around odd magnetic sites $1$ and $3$.
In the $E$-phase, the same symmetry operations can be applied in the planes $z= \pm$$c/2$
and also
will transform the
plane $z= c/2$ to the equivalent to it plane $z= -$$c/2$.
The magnetic texture in Fig.~\ref{fig.iEcanting}(a) is obtained by
the additional inversion around odd magnetic sites in the plane $z= c/2$, which
interchanges the symmetry operations $\{ \hat{C}^2_a| \boldsymbol{a}/2$$+$$\boldsymbol{b}/2 \}$ and
$\{ \hat{C}^2_a| -$$\boldsymbol{a}/2$$-$$\boldsymbol{b}/2 \}$. Thus, the plane
$z= c/2$ can be transformed to itself by the symmetry operation $\{ \hat{C}^2_a| \boldsymbol{a}/2$$+$$\boldsymbol{b}/2 \}$
around odd sites or by $\{ \hat{C}^2_a| -$$\boldsymbol{a}/2$$-$$\boldsymbol{b}/2 \}$
around even sites. Therefore, the symmetry operations $\{ \hat{C}^2_a| \boldsymbol{a}/2$$+$$\boldsymbol{b}/2 \}$
and $\{ \hat{C}^2_a| -$$\boldsymbol{a}/2$$-$$\boldsymbol{b}/2 \}$, although preserved locally in each of the plane,
are broken globally, because they cannot simultaneously transform the planes $z= 0$ and $z= \pm$$c/2$
to themselves.

  Instead, the magnetic texture in Fig.~\ref{fig.iEcanting}(a) obeys the symmetry operation
$\{ \hat{C}^2_c| \boldsymbol{c}/2\}$, which is another symmetry
operation of the space group $Pbnm$.
Therefore, the FE polarization in this phase will be parallel to the $\boldsymbol{c}$ axis.
According to the above arguments, each plane may carry a finite polarization parallel to the
$\boldsymbol{a}$ axis. However, since neighboring planes
are connected by the symmetry operation $\{ \hat{C}^2_c| \boldsymbol{c}/2\}$,
the contributions from different planes will cancel each other.

  The behavior of ${\bf P} || \boldsymbol{c}$,
obtained in the DE model for YMnO$_3$, is explained in Fig.~\ref{fig.iEcanting}(b). Here,
we again consider a continuous rotation of spins between two kinds of the AFM domains via an intermediate
spin-spiral phase, as explained in Fig.~\ref{fig.iEcanting}(c). In comparison with
Fig.~\ref{fig.Ecanting}, the planes $z = 0$ and $z = c/2$ are connected by the FM pathes between even magnetic sites.

  ${\bf P} || \boldsymbol{c}$ appears to be about two orders of magnitude weaker
than ${\bf P} || \boldsymbol{a}$ in the $E$-phase (Fig.~\ref{fig.Ecanting}).
Nevertheless, this result is very natural and can be easily understood by considering the
perturbation theory arguments, similar to the ones in Sec.~\ref{subsec:5orbitals}.
Namely, in order to obtain ${\bf P} || \boldsymbol{c}$, we should consider the transfer
integrals $\hat{\mathsf{t}}_{ij}$ between all possible combinations of sites $i$ and $j$ along the $\boldsymbol{c}$ axis.
Of course, the main contribution is
expected from the NN sites.
Moreover, according to Eq.~(\ref{eqn:PtotalE}), in order to contribute to ${\bf P} || \boldsymbol{c}$,
these transfer integrals
should have both symmetric and antisymmetric components.
However, due to the combination of $\{ \hat{C}^2_c| \boldsymbol{c}/2\}$
and the inversion symmetry around the Mn sites,
the NN integrals between the planes $z = 0$ and $z = c/2$ will satisfy the following property:
$\hat{R}^2_c \hat{t}_{ij} ( \hat{R}^2_c )^T = \hat{t}_{ji}$, where the matrix
transformation $\hat{R}^2_c$, corresponding to the $180^\circ$ rotation around the $\boldsymbol{c}$ axis,
changes the sign of some of the matrix elements of $\hat{t}_{ij}$. Therefore, in the crystal-field
representation, one can always choose the phases of the basis orbitals such that the corresponding
matrix of the transfer integrals $\hat{\mathsf{t}}_{ij}$ would become totally symmetric.
Thus, the NN contributions to ${\bf P} || \boldsymbol{c}$ in the second order of $1/\Delta$
will vanish, and ${\bf P} || \boldsymbol{c}$ has finite value due to either next-NN integrals, which are
small (all transfer integrals for YMnO$_3$ can be found in the Supplemental Material
of Ref.~\onlinecite{PRB12}) or the higher-order effects with respect to $1/\Delta$, which are also small.
This naturally explains the fact that ${\bf P} || \boldsymbol{c}$ is much smaller
than ${\bf P} || \boldsymbol{a}$.

  This finding resembles the behavior of multiferroic manganites with nearly fourfold periodic magnetic texture,
for which the possibility of
switching the electric polarization was demonstrated experimentally.\cite{Ishiwata,Kimura}
For example, in TbMnO$_3$
the polarization is aligned along the orthorhombic $\boldsymbol{c}$ axis.
However, the external magnetic field applied along the $\boldsymbol{b}$ axis will change
the magnetic texture and align the
polarization parallel to the $\boldsymbol{a}$ axis.\cite{Kimura}
Moreover, most of experimental data also confirm the fact that ${\bf P} || \boldsymbol{c}$ is smaller
than ${\bf P} || \boldsymbol{a}$.
For example, such a behavior is typical for the Eu$_{1-x}$Y$_x$MnO$_3$ compounds,
containing only nonmagnetic rare-earth elements, that excludes the influence
of the $4f$ magnetism
on the FE polarization.\cite{Ishiwata,Noda}
The results of this section suggest that this behavior is more generic and
can be anticipated in other regimes, including the twofold periodic magnetic systems.
The origin of this phenomenon is related to the specific symmetry of the crystal structure
(in the case of orthorhombic manganites -- the $Pbnm$ symmetry) and how it is lowered
by the magnetic ordering in the DE limit. It should not be confused with the spin-spiral
alignment, which does break
the inversion symmetry of the DE Hamiltonian (see Sec.~\ref{subsec:5orbitals}).

  In is interesting to note that the magnetic texture depicted in Fig.~\ref{fig.iEcanting}
can be viewed as a ``defected $E$-type AFM texture'', where the ``defects'' are
two FM bonds between the planes $z = 0$ and $z = c/2$.
Of course, such ``defects'' are energetically unfavorable and, after including the SO
interaction, this magnetic texture will change in order to minimize the FM coupling in the defected bonds.
This will lead to the substantial deformation of the magnetic texture in Fig.~\ref{fig.iEcanting}(a).
Nevertheless, we would like to emphasize that the noncollinear magnetic texture with
${\bf P} || \boldsymbol{c}$ can be stabilized even after including the SO interaction.
The situation was discussed in Ref.~\onlinecite{PRB11}.

\section{\label{sec:summary}  Discussions and Conclusions}

  This work is a continuation of previous studies,
devoted to multiferroic manganites, which crystallize in the orthorhombic
$Pbnm$ and $P2_1nm$ structure.\cite{PRB11,PRB12} Our main motivation
was to present a transparent physical picture, which would explain why and how the ferroelectric polarization
is induced by some complex magnetic order. For these purposes we invoke the double exchange theory,
which was formulated for the low-energy model, derived from the first-principles electronic structure
calculations. As far as the polarization is concerned, the DE theory is very robust and reproduces results of
more general mean-field Hartree-Fock calculations at a good quantitative level. Furthermore, the main
advantage of the DE
theory is that
it allows us to greatly simplify the problem and, in a number of cases, derive an analytical
expression for the FE polarization.
Thus, we could clarify very basic aspects of the FE activity in manganites with
twofold periodic magnetic texture.

  In our analysis we started from the general Berry-phase theory.\cite{KSV,Resta}
In the case of improper ferroelectrics, the basic quantity to be considered is the electronic polarization,
which incorporates the change of the electronic structure in response to the noncentrosymmetric alignment
of spins. Then, our main message is that, for the analysis of electronic polarization in realistic manganites,
one can always use two physical limits. The first one is the limit of large intra-atomic splitting $\Delta_{\rm ex}$
between the majority- and minority-spin states. The second one is the limit of large intra-atomic splitting $\Delta$
between the majority-spin $e_g$ states.
Therefore, for the electronic polarization, one can always consider the
perturbation theory expansion with respect to both $1/\Delta_{\rm ex}$ and $1/\Delta$.
This perturbation theory describes asymmetric transfer of some weight of the Wannier functions to the neighboring sites,
which gives rise to the polarization.

  There is some similarity with the theory of
superexchange interactions, which deals with the virtual hoppings,\cite{PWA}
and where the terms proportional to $1/\Delta$ and $1/\Delta_{\rm ex}$
account for the FM and AFM contributions, respectively.\cite{KugelKhomskii}
Therefore, the DE limit $\Delta_{\rm ex}$$\rightarrow$$\infty$
would correspond to neglecting all AFM contributions. It may not be a good approximation for
interatomic magnetic interactions.
Nevertheless, the main difference for the electronic polarization is that it appears only in the second order
with respect to $1/\Delta$ and $1/\Delta_{\rm ex}$.
The physically relevant picture corresponds to the situation where $\Delta_{\rm ex} > \Delta$.
Then, due to the inequality $(\Delta/\Delta_{\rm ex})^2 \ll \Delta/\Delta_{\rm ex}$,
it is logical to keep the effects of the first order of $1/\Delta_{\rm ex}$ in the
analysis of superexchange interactions,
but neglect the effects of the second order of $1/\Delta_{\rm ex}$ in the
analysis of electronic polarization.
This again justifies the use of the DE limit in the latter case.

  On the basis of this perturbation theory expansion, we
were able to explain how the electronic polarization depends on the
relative directions of spins in an arbitrary twofold periodic magnetic texture.
Particularly, the multiferroic effect in orthorhombic manganites is a nonlocal phenomenon in the sense that the
inversion symmetry is broken by making some of the Mn-Mn bonds magnetically inequivalent. In the DE model,
this inequivalence is achieved by the additional modulation of transfer integrals by $\xi_{ij}$. Then,
one trivial conclusion is that
there will be no magnetic inversion symmetry breaking in the spin-spiral phase, where all $\xi_{ij}$
are the same. Therefore, in order to make finite polarization, it is essential to deform the spin spiral.
In orthorhombic manganites, such deformation is caused by the relativistic spin-orbit interaction.\cite{PRB11,PRB12}
The second important precondition for the FE activity is the asymmetry of the transfer integrals, which should
simultaneously have symmetric and
antisymmetric components.

  We also pointed out on a serious problem in the structural optimization, which apparently exists in the
first-principles calculations (at least at the level of LDA$+$$U$ and GGA$+$$U$ approximations for the
exchange-correlation functional without relativistic spin-orbit coupling) and which typically
results in the large overestimation of
the value of FE polarization in comparison with experimental data.\cite{PRB12}
In this work, we were able to clarify the origin of this problem: in the theoretical structure, the directions
of noncentrosymmetric atomic displacements are inconsistent with the type of the orbital ordering
in the ferromagnetic zigzag chains, which controls the sign of the electronic polarization.
As the result, the electronic and ionic contributions have the same sign
in the theoretically optimized structure,
while, according to the experimental crystal structure, they should have opposite signs and partially cancel each other.

  Finally, we explained how the electronic polarization can be switched between
orthorhombic $\boldsymbol{a}$ and $\boldsymbol{c}$ directions by inverting the magnetic texture in every second
$\boldsymbol{ab}$ plane. We also expect a gigantic change of the absolute value of the polarization itself, which
is related to
very different symmetry properties of the nearest-neighbor transfer integrals
along the $\boldsymbol{c}$ direction
and in the $\boldsymbol{ab}$ plane of manganites.

  In this work, our analysis was limited by twofold periodic magnetic textures, which illustrate the
basic idea of the double exchange theory of ferroelectric polarization.
The idea can be extended to the systems with more general magnetic periodicity: apart from the
additional complexity of the magnetic texture, there is no fundamental difference between twofold
and more general magnetic periodicity. In both cases, the basic property, which should be considered
and which gives rise to the ferroelectric activity is the alternation of angles between spins in
different Mn-Mn bonds.

  \textit{Acknowledgements}.
  This work is partly supported by the grant of the Ministry of
Education and Science of Russia N 14.A18.21.0889.


\begin{thebibliography}{99}

\bibitem{MF_review}
Y. Tokura, Science  \textbf{312}, 1481 (2006);
T. Kimura, Annu. Rev. Mater. Res. \textbf{37}, 387 (2007);
S.-W. Cheong and M. Mostovoy,
Nature Materials \textbf{6}, 13 (2007);
D. Khomskii, Physics \textbf{2}, 20 (2009).

\bibitem{Ishiwata}
S. Ishiwata, Y. Kaneko, Y. Tokunaga, Y. Taguchi, T. Arima, and Y. Tokura,
Phys. Rev. B \textbf{81}, 100411 (2010).

\bibitem{Kimura}
T. Kimura, T. Goto, H. Shintani, K. Ishizaka, T. Arima, and Y. Tokura,
Nature \textbf{426}, 55 (2003);
T. Kimura, G. Lawes, T. Goto, Y. Tokura, and A.~P. Ramirez,
Phys. Rev. B \textbf{71}, 224425 (2005).

\bibitem{SergienkoPRL}
I.~A. Sergienko, C. \c Sen, and E. Dagotto,
Phys. Rev. Lett. \textbf{97}, 227204 (2006).

\bibitem{Picozzi}
S. Picozzi, K. Yamauchi, B. Sanyal, I.~A. Sergienko,
and E. Dagotto, Phys. Rev. Lett. \textbf{99}, 227201 (2007).

\bibitem{spiral_theories}
H. Katsura, N. Nagaosa, and A.~V. Balatsky,
Phys. Rev. Lett. \textbf{95}, 057205 (2005);
M. Mostovoy,
\textit{ibid.} \textbf{96}, 067601 (2006);
I.~A. Sergienko and E. Dagotto,
Phys. Rev. B \textbf{73}, 094434 (2006).

\bibitem{PRB11}
I.~V. Solovyev,
Phys. Rev. B \textbf{83}, 054404 (2011).

\bibitem{PRB12}
I.~V. Solovyev, M.~V. Valentyuk, and V.~V. Mazurenko,
Phys. Rev. B \textbf{86}, 144406 (2012).

\bibitem{Mochizuki}
M. Mochizuki, N. Furukawa, and N. Nagaosa,
Phys. Rev. Lett. \textbf{105}, 037205 (2010).

\bibitem{KSV}
R.~D. King-Smith and D. Vanderbilt,
Phys. Rev. B \textbf{47}, 1651 (1993);
D. Vanderbilt and R.~D. King-Smith, \textit{ibid.} \textbf{48}, 4442 (1993).

\bibitem{Resta}
R. Resta, J. Phys.: Condens. Matter \textbf{22}, 123201 (2010).

\bibitem{PWA}
P.~W.~Anderson,
Phys.~Rev.~{\bf 115}, 2 (1959).

\bibitem{KugelKhomskii}
K.~I. Kugel and D.~I. Khomskii,
Sov. Phys. Usp. \textbf{25}, 231 (1982).

\bibitem{DE_oldies}
C. Zener, Phys. Rev. \textbf{82} 440 (1951);
P. W. Anderson and H. Hasegawa, Phys. Rev. {\bf 100} 675 (1955);
P.-G. de Gennes, Phys. Rev. \textbf{118} 141 (1960);
K. Kubo and N. Ohata, J. Phys. Soc. Jpn. \textbf{33} 21 (1972).

\bibitem{review2008}
I.~V. Solovyev,
J. Phys.: Condens. Matter \textbf{20}, 293201 (2008).

\bibitem{JPSJ}
I. Solovyev,
J. Phys. Soc. Jpn. \textbf{78}, 054710 (2009).

\bibitem{Okuyama}
D. Okuyama, S. Ishiwata, Y. Takahashi, K. Yamauchi, S. Picozzi, K. Sugimoto, H. Sakai,
M. Takata, R. Shimano, Y. Taguchi, T. Arima, and Y. Tokura,
Phys. Rev. B \textbf{84}, 054440 (2011).

\bibitem{Yamauchi}
K. Yamauchi, F. Freimuth, S. Bl\"{u}gel, and S. Picozzi,
Phys. Rev. B \textbf{78}, 014403 (2008).

\bibitem{Barone}
P. Barone, K. Yamauchi, and S. Picozzi,
Phys. Rev. Lett. \textbf{106}, 077201 (2011).

\bibitem{PRB01}
I.~V. Solovyev and K. Terakura,
Phys. Rev. Lett. \textbf{83}, 2825 (1999);
I.~V. Solovyev,
Phys. Rev. B \textbf{63}, 174406 (2001).

\bibitem{Kanamori}
J. Kanamori, J. Appl. Phys. \textbf{31}, 14S (1960).

\bibitem{SK}
J.~C. Slater and G.~F. Koster,
Phys. Rev. \textbf{94}, 1498 (1954).

\bibitem{Zhang}
X.~L. Qi, T.~L. Hughes, and S.-C. Zhang,
Phys. Rev. B \textbf{78}, 195424 (2008).

\bibitem{Hotta}
T. Hotta, M. Moraghebi, A. Feiguin, A. Moreo, S. Yunoki, and E. Dagotto,
Phys. Rev. Lett. \textbf{90}, 247203 (2003).

\bibitem{remark1}
For $\Delta$$=$$0$, the integral (\ref{eqn:Pfinal}) can be evaluated analytically:
note that $\int_0^{2\pi/a}dk = \int_0^{\pi/a}dk + \int_{\pi/a}^{2\pi/a}dk$, and
use that $\int_{\pi/a}^{2\pi/a}dk$ can be further transformed to $\int_0^{\pi/a}dk$ after replacing
$d_z(k)$ by $-$$d_z(k)$.

\bibitem{Zak}
J. Zak, Phys. Rev. Lett. \textbf{62}, 2747 (1989).

\bibitem{Munoz}
A. Mu\~noz, M. T. Cas\'ais, J. A. Alonso, M. J. Mart\'inez-Lope,
J. L. Mart\'inez, and M. T. Fern\'andez-D\'iaz,
Inorg. Chem. \textbf{40}, 1020 (2001).

\bibitem{HeFranchini}
J. He and C. Franchini,
Phys. Rev. B \textbf{86}, 235117 (2012).

\bibitem{Stroppa}
A. Stroppa and S. Picozzi,
Phys. Chem. Chem. Phys. \textbf{12}, 5405 (2010).

\bibitem{Malashevich}
A. Malashevich and D. Vanderbilt,
Phys. Rev. Lett. \textbf{101}, 037210 (2008);
Phys. Rev. B \textbf{80}, 224407 (2009).

\bibitem{Noda}
K. Noda, M. Akaki, F. Nakamura, D. Akahoshi, and H. Kuwahara,
J. Magn. Magn. Matter. \textbf{310}, 1162 (2007).


\end{thebibliography}
\end{document}